\documentclass[pra,twocolumn,amsmath,amssymb,superscriptaddress,longbibliography]{revtex4-1}
\usepackage{graphicx,amsmath,relsize,epstopdf,upgreek,color,mathtools,bm,mathptmx}
\usepackage[colorlinks=true,linkcolor=blue,citecolor=blue,urlcolor =blue]{hyperref}

\newcommand{\sugg}[1]{\textcolor{black}{#1}}
\newcommand{\suggtwo}[1]{\textcolor{black}{#1}}
\newcommand{\us}{\textscale{1.5}{\textunderscore}}

\newcommand{\ket}[1]{\left|{#1}\right>}
\newcommand{\bra}[1]{\left<{#1}\right|}

\newcommand{\opinner}[3]{\left<{#1}\vphantom{#1}\vphantom{#3}\right|{#2}\left|{#3}\vphantom{#1}\vphantom{#3}\right>}
\newcommand{\rvec}[1]{\pmb{#1}}
\newcommand{\dyadic}[1]{\pmb{#1}}

\newcommand{\D}{\mathrm{d}}
\newcommand{\I}{\mathrm{i}}
\newcommand{\TP}[1]{{#1}^\mathrm{\,\textsc{t}}}
\newcommand{\E}[1]{\mathrm{e}^{\mbox{\footnotesize$#1$}}}

\newcommand{\DET}[1]{\det\!\left\{#1\right\}}

\newcommand{\ML}[1]{\widehat{#1}_\textsc{ml}}

\newcommand{\FML}{\dyadic{F}_\textsc{ml}}
\newcommand{\invFML}{\dyadic{F}^{-1}_\textsc{ml}}
\newcommand{\gML}{\bm{g}_\textsc{ml}}
\newcommand{\BETA}[1]{\mathrm{B}\!\left(#1\right)}

\newcommand{\tr}[1]{\mathrm{tr}\!\left\{#1\right\}}
\newcommand{\Tr}[1]{\mathrm{Tr}\!\left\{#1\right\}}
\newcommand{\cmpl}{\mathrm{cmpl}}

\catcode`,\active

\catcode`\,12

\allowdisplaybreaks
\begin{document}

\title{Efficient Bayesian credible-region certification for quantum-state tomography}

\author{C.~Oh}
\affiliation{Center for Macroscopic Quantum Control, Department of Physics and Astronomy, Seoul National University, 08826 Seoul, South Korea}

\author{Y.~S.~Teo}
\email{ys\_teo@snu.ac.kr}
\affiliation{Center for Macroscopic Quantum Control, Department of Physics and Astronomy, Seoul National University, 08826 Seoul, South Korea}
\affiliation{Frontier Physics Research Division, Department of Physics and Astronomy, Seoul National University, 08826 Seoul, South Korea}

\author{H.~Jeong}
\affiliation{Center for Macroscopic Quantum Control, Department of Physics and Astronomy, Seoul National University, 08826 Seoul, South Korea}

      \begin{abstract}
      	Standard Bayesian credible-region theory for constructing an error region on the unique estimator of an unknown state in general quantum-state tomography to calculate its size and credibility relies on heavy Monte~Carlo sampling of the state space followed by \sugg{filtering to obtain the correct region sample}. This conventional \sugg{methodology} typically gives negligible yield for very small error regions originating from large datasets. \sugg{In this article, we discuss at length the in-region sampling theory for computing} both size and credibility from region-average quantities that \sugg{avoids this general problem altogether}. \sugg{Among the many possible numerical choices, we study the performance and properties of} accelerated hit-and-run Monte~Carlo \sugg{algorithm for in-region sampling} and provide its complexity estimates for quantum states. Finally with our in-region concept, \sugg{by alternatively quantifying the region capacity with} the region-average distance between two states in the region (measured for instance with either the Hilbert-Schmidt, trace-class or Bures distance), we derive approximation formulas to analytically estimate both \sugg{region capacity} and credibility without Monte~Carlo computation.
      \end{abstract}

\maketitle

\section{Introduction}

All physical-quantity estimates obtained from collected data should be accompanied by ``error-bars'' to accurately convey all properties of the physical system of interest. This applies to quantum-state tomography~\cite{Smithey:1993uq,Chuang:2000fk,Rehacek:2007ml,Teo:2011me,Zhu:2014aa}, which is an important preliminary step for implementing all quantum cryptography and computation protocols~\cite{Ladd:2010aa,Campbell:2017aa,Ladd:2010aa,Lekitsch:2017aa} reliably.

Bootstrapping procedures~\cite{Efron:1993bs,Davison:1997ri} are amongst some of the most widely-used techniques for assigning ``error-bars'' to reconstructed quantum states. Recently, it was pointed out in~\cite{Suess:2017np} that such assignments lack rigorous statistical foundations and may produce ``error-bars'' that are too small for reliable conclusions. The rather more justified approach falls under the study of hypothesis testing~\cite{Christensen:2005ht}. Two grand schools of thought exist for this purpose. In the context of quantum-state reconstruction of an unknown state $\rho$, one may treat $\rho$ as ``absolute'' (the frequentist school) and attempt to extract this knowledge from collected data. This suggests the constructions of \emph{confidence regions}~\cite{Christandl:2012qs,Blume-Kohout:2012eb,Faist:2016eb}, which are error-regions for the state estimator $\widehat{\rho}$ from all plausible datasets, including those unmeasured in the experiment. An accurate $\widehat{\rho}$ for $\rho$ would then entail a collection of typically small confidence regions with high probability that $\rho$ for each plausible dataset lies in the corresponding region.

Given that only one dataset (the measured one) is really available to the observer, we shall focus on the apt Bayesian school of thought that instead regards this dataset as ``absolute'' and constructs \emph{credible regions}~\cite{Shang:2013cc,Li:2016da} as the error regions for $\widehat{\rho}$ beginning with some prior distribution $p(\rho)$ of $\rho$. A fairly accurate estimator $\widehat{\rho}$ for some unknown $\rho$ naturally implies a credible region (generated from the measured dataset) of a small size with a large probability that this $\rho$ is inside the region---a high credibility~\cite{Teo:2018aa}. In order to obtain a reasonably small error region (be it that of credible- or confidence-type), one may either resort to adaptive strategies~\cite{Oh:2018aa} and optimize additional properties of the region, or simply increase the dataset collected in quantum tomography.

The complicated quantum state-space boundary~\cite{Bengtsson:2006gm,Bengtsson:2013gm} renders any analytical attempt at calculating size and credibility for any credible region futile, leaving numerical Monte~Carlo (MC) methods~\cite{Shang:2015mc,Seah:2015mc} as the only viable option. As the size of the credible region is defined as its volume fraction with the quantum state space, one needs an extremely large sample of the state space to finally end up having a reasonable sample for the region. Despite the optimistic advantages that some of these MC schemes may have in generating samples of arbitrary distributions, one deleterious issue for such an \sugg{MC-filtering strategy} becomes apparent when the dataset is large, which is the common situation in any tomography experiment. The resulting credible region eventually becomes too tiny relative to the quantum state space for \sugg{MC-filtering} to produce any effective yield to properly compute the size and credibility.

\sugg{To solve this problem, we introduced the in-region sampling technique~\cite{PRL} to feasibly compute all credible-region properties by simply sampling an appropriate quantity over the region itself. This follows from the logic that a change in the region-average quantity encodes the change in both region size and credibility. In our theory}, we prove the central lemma stating that the size (and credibility) of any credible region are related to a class of region-average quantities through a first-order differential equation that is solvable numerically. \sugg{As an example study, we discuss the computation of region-average quantities using the accelerated hit-and-run algorithm, its correlation properties,} and estimate its complexity from \sugg{geometrical considerations} of the quantum state space. The region-average formalism encourages \sugg{the formulation of the region capacity (a different way of stating ``how big'' a region is) by investigating} the average relative distance between two points in the region. This region-average distance may be induced by any of the common measures used in quantum information, and we shall explicitly consider the Hilbert-Schmidt, trace-class, and Bures distance measures as popular examples. It turns out that this perspective\sugg{, together with the results in \cite{Teo:2018aa}, offers closed-form analytical approximation formulas for an alternative rapid approximate Bayesian error certification with no Monte~Carlo methods necessary.}

This article is organized as follows: A preliminary introduction to the basic notions of standard Bayesian credible-region theory shall ensue in Sec.~\ref{sec:std_th}, and the stage for discussions with large data is set in Sec.~\ref{sec:bigdata}. Next, we present our \sugg{in-region sampling} theory for size and credibility that works for any kind of data and prior in Sec.~\ref{sec:newtheory}. Afterwards, we describe how region-average quantities can be numerically computed and estimate the computational complexities in Sec.~\ref{sec:numcomp}. Section~\ref{sec:newsize} then proceeds to \sugg{quantify the region capacity} in terms of region-average distances induced by all the three aforementioned distance measures \sugg{other than the Bayesian size}. Finally, for fast analytical Bayesian error estimates, we derive asymptotic formulas for all important region-average quantities in Sec.~\ref{sec:formulas} based on the perspective of distance-induced size. \sugg{All numerical results and computation correlation properties are then presented and discussed in Sec.~\ref{sec:results} with explicit examples in quantum tomography.}

\section{Standard Bayesian credible-region theory}
\label{sec:std_th}

Before a quantum-state tomography experiment commences, the observer might have some (justifiable) preconception about the unknown quantum state $\rho\geq0$ ($\tr{\rho}=1$) of Hilbert-space dimension $D$. Such preconception is usually not uniquely privileged, and therefore weighted with some \emph{prior probability distribution} $p(\rho)$. After the experiment, the observer collects a set of data $\mathbb{D}$ that are \emph{informationally complete} (IC) such that a unique estimator $\widehat{\rho}$ for $\rho$ is acquired. In quantum theory, the measurements are modeled as a \emph{probability-operator measurement} (POM) consisting \suggtwo{of} a set of $M$ positive operators $\Pi_j\geq0$ that sum to the identity. Associated to every such experiment is the \emph{likelihood function}~$L=L(\mathbb{D}|\widehat{\rho})$, with which the observer obtains a \emph{posterior probability distribution} (knowledge after-the-fact) that is a function of $L$.

It was formerly established in~\cite{Shang:2013cc} that for this measured dataset $\mathbb{D}$, if $\widehat{\rho}$ is taken to be the estimator that maximizes $L$---the maximum-likelihood (ML) estimator---, then a Bayesian credible region (CR) $\mathcal{R}$ can be constructed around $\ML{\rho}$, which turns out to have a constant likelihood boundary $\partial{\mathcal{R}}$ within the quantum state space $\mathcal{R}_0$. For this CR, which is a subregion of $\mathcal{R}_0$, we can specify its \emph{size} and \emph{credibility}, the latter which is the probability that $\rho\in\mathcal{R}$. Such a region is optimal in the sense that it gives the largest credibility for a given size, or equivalently possesses the smallest size for a given credibility.

In this article, we shall be interested in reconstructing the $(d=D^2-1)$-dimensional real vectorial parameter $\rvec{r}\leftrightarrow\rho$ that characterizes $\rho$. More technically, this equivalent parametrization is achieved with a Hermitian operator basis $\{1/\sqrt{D},\Omega_j\}^{d}_{j=1}$ that contains $d$ trace-orthonormal traceless operators $\Omega_j$ \sugg{$\left(\tr{\Omega_j\Omega_k}=\delta_{j,k}\right)$}, by which $\rvec{r}=\tr{\rho\,\rvec{\Omega}}$ is defined from the column $\rvec{\Omega}$ of $\Omega_j$s.
Formally, in terms of the multivariate parameter $\rvec{r}$, the size and credibility of $\mathcal{R}=\mathcal{R}_\lambda$ for some $0\leq\lambda\leq1$ are respectively given by~\cite{Shang:2013cc}
\begin{align}
S_\lambda\equiv&\,\int_{\mathcal{R}_\lambda}(\D\,\rvec{r}')=\int_{\mathcal{R}_0}(\D\,\rvec{r}')\,\eta(L-\lambda L_\text{max})\,,\nonumber\\
C_\lambda\equiv&\,\dfrac{1}{L(\mathbb{D})}\int_{\mathcal{R}_\lambda}(\D\,\rvec{r}')\,L=\dfrac{1}{L(\mathbb{D})}\int_{\mathcal{R}_0}(\D\,\rvec{r}')\,\eta(L-\lambda L_\text{max})\,L\,,
\label{eq:std_defs}
\end{align}
where the volume measure $(\D\,{\rvec{r}})$ incorporates the prior distribution $p(\rvec{r})$ for $\rvec{r}$, $\eta$ is the Heaviside function, $L(\mathbb{D})=\int_{\mathcal{R}_0}(\D\,\rvec{r}')\,L(\mathbb{D}|\rvec{r}')$. The important variable $0\leq\lambda\leq1$ specifies the shape and size of $\mathcal{R}_\lambda$, from which the limits $\mathcal{R}_{\lambda=0}=\mathcal{R}_0$ and $\mathcal{R}_{\lambda=1}=\{\ML{\rvec{r}}\}$ are immediate. Here, we note that the probability parametrization was adopted in \cite{Shang:2013cc}. Upon the condition that each datum measurement, corresponding to an outcome $\Pi_j$, the inherent statistics of $\mathbb{D}$ is therefore multinomial and the log-likelihood reads $\log L=\sum_jn_j\log p'_j$ with the collected relative frequencies $\sum_jn_j=N$ that make up $N$ measured data copies, and $p'_j=\tr{\rho'\Pi_j}$ for any state $\rho'$.

\begin{figure}[t]
	\center
	\includegraphics[width=1\columnwidth]{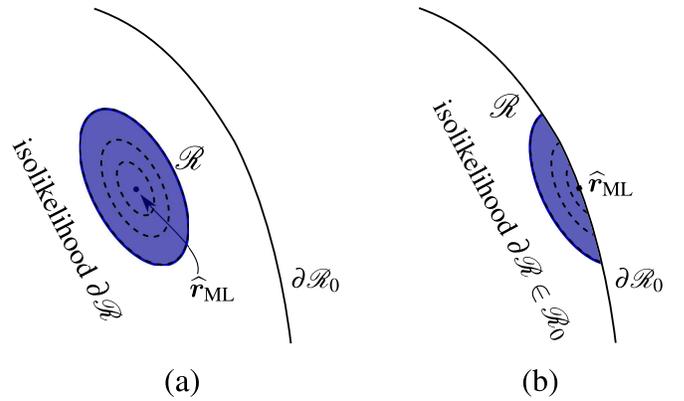}
	\caption{\label{fig:caseAB}(Color Online) Credible region $\mathcal{R}$ in (a)~Case~A and (b)~Case~B, consolidating the two very general situations that could happen in the limit $N\gg1$~\sugg{\cite{PRL}}. Case~A corresponds to a state estimator $\ML{\rho}\leftrightarrow\ML{\rvec{r}}$ that is full-rank, whereas Case~B implies that the estimator is rank-deficient.}
\end{figure}

We can gain a clear physical picture of both size and credibility: they respectively quantify the prior and posterior content $\mathcal{R}$, hence the symbol $S_\lambda$ for the former. Owing to the dual nature of size $S_\lambda$ and credibility $C_\lambda$, it can be shown, in fact, that 
\begin{equation}
C_\lambda=\dfrac{\displaystyle\lambda S_\lambda+\int^1_\lambda\D\,\lambda' S_{\lambda'}}{\displaystyle\int^1_0\D\,\lambda' S_{\lambda'}}\,.
\label{eq:Cpr}
\end{equation}
Put differently, $C_\lambda$ may be straightforwardly computable through single-parameter integrations in $\lambda$ so long as $S_\lambda$ is known up to some arbitrary constant multiple.

Nonetheless, the complicated boundary $\partial\mathcal{R}_0$ of the quantum state space makes the computation of $S_\lambda$ extremely challenging even numerically. The innate definition of $S_\lambda$, namely the volume fraction of $\mathcal{R}_\lambda$ to $\mathcal{R}_0$, requires, first, the generation of a sufficiently large sample of $\mathcal{R}_0$, followed by the \sugg{filtering} of all its sampled points that lie \sugg{inside} $\mathcal{R}_\lambda$ for any $\lambda$. There exist various Monte~Carlo (MC) methods to sample $\mathcal{R}_0$~\cite{Shang:2015mc}. Ultimately, this \sugg{MC-filtering} strategy exhibits one major disadvantage: in the limit of large data sample ($N\gg1$), $\mathcal{R}_\lambda$ would become so small relative to $\mathcal{R}_0$ that the \sugg{MC-filtering} strategy needs a sufficiently large number of random MC sample points from $\mathcal{R}_0$ to produce any useful yield. The scaling of MC sample size needed to maintain a fixed yield, which was estimated to be $O(N^{d/2})$~\cite{Teo:2018aa}, thus outgrows the feasible computational yield-rate very quickly. The bottom-line: a much more feasible numerical strategy to perform Bayesian error certification is necessary in this practical data limit.

\section{The large-data condition}
\label{sec:bigdata}

Before presenting an alternative operational theory, unless otherwise stated, we shall consider $N\gg1$ as the putative limit in pragmatic tomography experiments. We emphasize here that $N$ only has to be sufficiently large for the statistical central limit theorem to dictate a Gaussian form for $L$. In this limit, there can only be one of two cases: either $\mathcal{R}$ is completely inside $\mathcal{R}_0$ that contains a full-rank $\ML{\rho}$ (Case~A) or partially truncated by the state-space boundary $\partial\mathcal{R}_0$ of $\mathcal{R}_0$ that houses a rank-deficient $\ML{\rho}$ (Case~B) (see Fig.~\ref{fig:caseAB}). 

Case~A arises when the unknown state $\rho$ is away from $\partial\mathcal{R}_0$, so that a sufficiently large $N$ would produce untruncated regions for $\lambda$ values corresponding to desirably large $C_\lambda<1$. This case offers a simple geometrical description for $\mathcal{R}$. Upon invoking the Taylor expansion of 
\begin{equation}
\log L(\mathbb{D}|\rvec{r}')\approx\log L_\text{max}-\dfrac{1}{2}(\rvec{r}'-\ML{\rvec{r}})\bm{\cdot}\FML\bm{\cdot}(\rvec{r}'-\ML{\rvec{r}})
\end{equation}
about the interior $\ML{\rvec{r}}$ up to the second order, we find that the likelihood $L$ is essentially a Gaussian function centered at $\ML{\rvec{r}}$ of height $L_\text{max}$, with its covariance profile shaped by $\FML$, that is the \emph{Fisher information} evaluated at $\ML{\rvec{r}}$. The CRs $\mathcal{R}_\lambda$ that go with this Gaussian likelihood are, hence, simple hyperellipsoids $\mathcal{E}_\lambda$ described by the inequality $(\rvec{r}'-\ML{\rvec{r}})\bm{\cdot}\FML\bm{\cdot}(\rvec{r}'-\ML{\rvec{r}})\leq-2\log\lambda$.

If $\rvec{r}$ is located in $\partial\mathcal{R}_0$, then as $N$ increases, the ML estimator $\ML{\rvec{r}}$ would eventually approach $\rvec{r}$ and there is a high probability that $\ML{\rvec{r}}\in\partial\mathcal{R}_0$ before this happens. For sufficiently large $N$, we have Case~B where $\partial\mathcal{R}\cap\partial\mathcal{R}_0$ is not disjointed and falls on the side of $\ML{\rvec{r}}$. To asymptotically cope with such a situation, we may again expand
\begin{align}
\log L(\mathbb{D}|\rvec{r}')\approx&\,\log L_\text{max}+(\rvec{r}'-\ML{\rvec{r}})\bm{\cdot}\gML\nonumber\\
&\,-\dfrac{1}{2}(\rvec{r}'-\ML{\rvec{r}})\bm{\cdot}\FML\bm{\cdot}(\rvec{r}'-\ML{\rvec{r}})\nonumber\\
=&\,\log L'_\text{max}-\dfrac{1}{2}(\rvec{r}'-\rvec{r}_\text{c})\bm{\cdot}\FML\bm{\cdot}(\rvec{r}'-\rvec{r}_\text{c})
\label{eq:caseB_clt}
\end{align}
about the boundary $\ML{\rvec{r}}$, where this time $L$ is a Gaussian function centered at $\rvec{r}_\text{c}=\ML{\rvec{r}}+\FML^{-1}\bm{\cdot}\gML$ with $\gML=\partial\log L(\mathbb{D}|\rvec{r}')/\partial\rvec{r}'|_{\rvec{r}'=\ML{\rvec{r}}}$, and possesses a height $L'_\text{max}=L_\text{max}\,\exp(\gML\bm{\cdot}\FML^{-1}\bm{\cdot}\gML/2)>L_\text{max}$. The covariance profile of this Gaussian function is still governed by $\FML$, which produces hyperellipsoids $\mathcal{E}'_\lambda$ described according to $(\rvec{r}'-\rvec{r}_\text{c})\bm{\cdot}\FML\bm{\cdot}(\rvec{r}'-\rvec{r}_\text{c})\leq-2\log\lambda'$ for an ``effective $\lambda$'' $\lambda'$ defined by $2\log(\lambda/\lambda')=\gML^\textsc{t}\FML^{-1}\gML$. The CR $\mathcal{R}_\lambda$ is then asymptotically $\mathcal{E}'_\lambda\cap\mathcal{R}_0$.

We point out that there is an intermediate case in which $\mathcal{R}_\lambda=\mathcal{E}_\lambda$, centered at $\ML{\rvec{r}}\notin\partial\mathcal{R}_0$, is truncated by $\partial\mathcal{R}_0$. Such a situation can happen when $N$ is not sufficiently large, and tends to either Case~A or B as $N$ grows. On a separate note, Ref.~\cite{Teo:2018aa,Oh:2018aa} explicitly studies also this intermediate case.

\begin{figure}[t]
	\center
	\includegraphics[width=1\columnwidth]{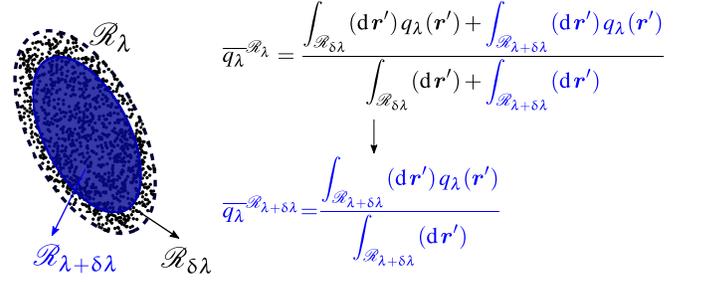}
	\caption{\label{fig:R-average}(Color Online) An infinitesimal change in $\lambda$ causes a transformation in $\mathcal{R}_\lambda$ that subsequently exclude all points in the hyperannulus $\mathcal{R}_{\updelta\lambda}$ from the region-average $\overline{q_\lambda}^{\mathcal{R}_{\lambda+\updelta\lambda}}$.}
\end{figure}

\section{In-region sampling theory}
\label{sec:newtheory}

Suppose we have a CR $\mathcal{R}_\lambda$, with which we define the average quantity
\begin{equation}
u_\lambda\equiv\overline{q_\lambda (\rvec{r}')}^{\mathcal{R}_\lambda}=\dfrac{\displaystyle\int_{\mathcal{R}_\lambda}(\D\,\rvec{r}')\,q_\lambda(\rvec{r}')}{\displaystyle\int_{\mathcal{R}_\lambda}(\D\,\rvec{r}')}=\dfrac{1}{K_\text{smp}}\sum^{K_\text{smp}}_{l=1}q_{\lambda,l}
\label{eq:ulbd}
\end{equation}
for some function $q_\lambda$, which is approximately equivalent to the discrete-sum average of $q_{\lambda,l}$ values over a sufficiently large number $K_\text{smp}$ of region points. If we probe the response of $u_\lambda$ with an incremental change $\lambda\rightarrow\lambda+\updelta\lambda$ in $\lambda$ as in Fig.~\ref{fig:R-average}, the result is the total change
\begin{equation}
\updelta u_\lambda=\left(\dfrac{1}{S_\lambda}-\dfrac{\updelta S_\lambda}{S^2_\lambda}\right)\int_{\mathcal{R}_{\updelta\lambda}}(\D\,\rvec{r}')\,q_\lambda-\dfrac{\updelta S_\lambda}{S_\lambda^2}\int_{\mathcal{R}_{\lambda}}(\D\,\rvec{r}')\,q_\lambda
\end{equation}
after limiting all small changes to the first order, which reveals that a small increment $\updelta u_\lambda$ can be explained by a change $\updelta S_\lambda$ in size that is accompanied by the (in)exclusion of the annular sum $\int_{\mathcal{R}_{\updelta\lambda}}(\D\,\rvec{r}')\,q_\lambda$. Put simply, tracking the change in $u_\lambda$ allows us to infer how much $S_\lambda$ has changed.

To better utilize this intuition, we first take the derivative of $u_\lambda S_\lambda$, which gives
\begin{align}
\dfrac{\partial u_\lambda S_\lambda}{\partial\lambda}=&\,-L_\text{max}\int_{\mathcal{R}_0}(\D\,\rvec{r}')\,\delta(L-\lambda\,L_\text{max})\,q_\lambda(\rvec{r}')\nonumber\\
+&\,\int_{\mathcal{R}_\lambda}(\D\,\rvec{r}')\,\dfrac{\partial q_\lambda(\rvec{r}')}{\partial\lambda}
\label{eq:lemma_inter}
\end{align}
after invoking the derivative identity $\D\eta(x)/\D x=\delta(x)$ between $\eta(x)$ and the Dirac delta function $\delta(x)$. \sugg{Next, we may impose the following functional form $q_\lambda(\rvec{r}')\equiv f(L)-f(\lambda L_\text{max})$ for $q_\lambda$, where $f(L)$ is some arbitrary function of $L$. This simplifies Eq.~\eqref{eq:lemma_inter} to
\begin{equation}
\dfrac{\partial}{\partial\lambda}(u_\lambda\,S_\lambda)=-S_\lambda\,\dfrac{\partial}{\partial\lambda}\,f(\lambda L_\text{max})\,.
\label{eq:ode}
\end{equation}}

We now have a first-order differential equation that describes the dynamics of $S_\lambda$ according to the parametric region-average $u_\lambda$. With the initial condition $S_{\lambda=0}=1$, the entire functional form of $S_\lambda$ can then be recovered with Eq.~\eqref{eq:ode}. This completes the constructive proof of our so-called\\[1ex]

\noindent
{\bf Region-average computation (RAC) lemma}: \emph{For any prior $(\D\,\rvec{r}')$ and measurement data $\mathbb{D}$, the prior content $S_\lambda$ (up to a multiplicative factor), and hence the credibility $C_\lambda$, are all inferable from $u_\lambda$ defined in Eq.~\eqref{eq:ulbd} with \sugg{$q_\lambda(\rvec{r}')\equiv f(L)-f(\lambda L_\text{max})$}.}\\[1ex]

\noindent
To proceed, we first perform the substitution \sugg{$y_\lambda=u_\lambda\,S_\lambda$} to yield another differential equation\sugg{
\begin{equation}
\dfrac{\partial y_\lambda}{\partial\lambda}=-\dfrac{y_\lambda}{u_\lambda}\,\dfrac{\partial}{\partial\lambda}\,f(\lambda L_\text{max})\,.
\label{eq:ode2}
\end{equation}}
The solution to $y_\lambda$ can then be obtained numerically through Euler's method~\cite{Butcher:2003aa}. In practice, we may start from \sugg{$S_{\lambda\approx0}\equiv1$} and iterate\sugg{
\begin{equation} y_{\lambda_{j+1}}=y_{\lambda_j}-\dfrac{y_{\lambda_j}}{u_{\lambda_j}}\,\dfrac{\partial}{\partial\lambda_j}\,f(\lambda_j L_\text{max})
\label{eq:euler}
\end{equation}}
for a sequence of discretized $\lambda\rightarrow\lambda_j$ values ranging from 0 to 1. For feasible computation of $u_\lambda$, we shall choose \sugg{$f(L)=\log L$}.

\section{Region-average numerical computation}
\label{sec:numcomp}

\subsection{The hit-and-run algorithm}

The hit-and-run algorithm is a direct convex-body MC sampling scheme that generates random sample points in the body according to some predefined distribution. This algorithm is thus suited for sampling $\mathcal{R}_\lambda$ according to some prior distribution $p(\rvec{r})$ for the unknown $\rvec{r}$.

The sampling principles behind an efficient hit-and-run computation begin with defining the smallest possible convex set $\mathcal{B}\supseteq\mathcal{R}_\lambda$ that houses $\mathcal{R}_\lambda$ and has an easy-access geometry. Starting from a known point in $\mathcal{R}_\lambda$, say the ML estimator $\ML{\rvec{r}}$, a random line segment passing through this point is generated, with its endpoints fixed at $\partial\mathcal{B}$ that are quickly computable because of its simple geometry. Following which, sampling commences by repeatedly picking a random point along this segment until it lies in $\mathcal{R}_\lambda$. This point is next taken to be the new reference point through which another line segment is generated to find a new random point in $\mathcal{R}_\lambda$, until a set of $K_\text{smp}$ points is gathered.

We can make use of the straightforward hyperellipsoidal characteristics inherent from the central limit theorem to construct $\mathcal{B}$. For Case~A, where $\mathcal{R}_\lambda=\mathcal{E}_\lambda$, $\mathcal{B}$ can just be taken to be $\mathcal{E}_\lambda$ characterized by $\widetilde{\FML}=\FML/(-2\log\lambda)$ from the earlier discussions in Sec.~\ref{sec:bigdata}. We now turn to the more interesting and practically ubiquitous Case~B, where the large-$N$ arguments of Sec.~\ref{sec:bigdata} imply that we may fix $\mathcal{B}=\mathcal{E}'_\lambda$, the profile of which is governed by $\widetilde{\FML'}=\FML/(-2\log\lambda')$. For this case, if $\mathcal{B}$ is much larger than $\mathcal{R}_\lambda$, sampling the latter would incur a significant amount of wastage. Fortunately there exists an accelerated version of hit-and-run~\cite{Kiatsupaibul:2011aa} that adaptively shrinks the endpoints of the line segment to reduce the search space each step.

To check if the random point chosen from the line segment is in $\mathcal{R}_\lambda$, we recall that $\mathcal{R}_\lambda$ is equivalently the continuous set of unit-trace operators that are both positive and satisfy the hyperellipsoidal constraint defined by the inequality $L(\mathbb{D}|\rvec{r}')>\lambda\,L_\text{max}$ under the $N\gg1$ limit. Since all points in $\mathcal{B}$ essentially fulfill the latter constraint, the primary task is to check if the random point correspond to a legitimate quantum state. It is known that the Cholesky decomposition~\cite{Horn:1985aa,Higham:2009aa}, a routine that factorizes $\rho'=A^\dagger A$ for any positive operator $\rho'$, is an efficient and numerically stable way for the job, with a computational complexity of $O(D^3)$. A routine failure implies that the operator corresponding to the selected point is not positive.

The complete pseudo-coded algorithm for the accelerated hit-and-run, \suggtwo{tailored for} an arbitrary prior distribution $p(\rvec{r})$, is stated below~\cite{Belisle:1993aa,Smith:1996aa,Kiatsupaibul:2011aa}:

\begin{center}
	\begin{minipage}[c][11cm][c]{0.9\columnwidth}
		\vspace{1ex}
		\noindent
		\rule{\columnwidth}{1.5pt}\\
		\textbf{Accelerated hit-and-run for \sugg{sampling $\mathcal{R}_\lambda$}}\\[1ex]
		\noindent
		Beginning with $k=1$ and $\rvec{r}_\text{ref}=\ML{\rvec{r}}$ of $N\gg1$:
		\begin{enumerate}
			\item Generate a random line segment characterized by $\rvec{y}=\rvec{r}_\text{ref}+\mu\,\rvec{\mathrm{e}_v}$, where $\rvec{\mathrm{e}_v}=\rvec{v}/|\rvec{v}|$ and $\rvec{v}$ follows the standard Gaussian distribution (mean 0 and variance 1 for each column entry). Its endpoints are parametrized by $\mu_\pm=[-b\pm\sqrt{b^2-a(c-1)}]/a$, where $\rvec{\Delta}=\rvec{r}_\text{ref}-\rvec{r}_\text{c}$, $a=\TP{\rvec{\mathrm{e}_v}}\,\dyadic{A}\,\rvec{\mathrm{e}_v}$, $b=\TP{\rvec{\Delta}}\,\dyadic{A}\,\rvec{\mathrm{e}_v}$, $c=\TP{\rvec{\Delta}}\,\dyadic{A}\,\rvec{\Delta}$, and $\dyadic{A}=\widetilde{\FML}\,\,\text{or}\,\,\widetilde{\FML}'$.
			\item Define $\beta_1\equiv\mu_\text{min}=\min\{\mu_+,\mu_-\}$ and $\beta_2\equiv\mu_\text{max}=\max\{\mu_+,\mu_-\}$.
			\item \sugg{Pick a random number $\beta_1\leq\beta\leq\beta_2$ according to the marginal probability distribution $p(\rvec{r}_\text{ref}+\beta\,\rvec{\mathrm{e}_v})/\int\D\beta'p(\rvec{r}_\text{ref}+\beta'\,\rvec{\mathrm{e}_v})$ truncated in the interval [$\beta_1,\beta_2$] and obtain $\rvec{r}_\text{test}=\rvec{r}_\text{ref}+\beta\,\rvec{\mathrm{e}_v}$.}
			\item Check whether $\rho_\text{test}\leftrightarrow\rvec{r}_\text{test}$ is positive.
			\begin{itemize}
				\item If so, define $\rvec{r}_\text{ref}=\rvec{r}_\text{test}$, raise $k$ by 1, and go to step~1. 
				\item If not, set $\beta_1=\beta$ if $\beta<0$, or $\beta_2=\beta$ if $\beta>0$, and repeat steps~3 and 4. 
			\end{itemize}
			\item End routine if $k>K_\text{smp}$, the total number of sample points desired.\\[-5.5ex]
		\end{enumerate}
		\rule{\columnwidth}{1.5pt}
		\vspace{1ex}
	\end{minipage}
\end{center}

\begin{figure}[t]
	\center
	\includegraphics[width=0.8\columnwidth]{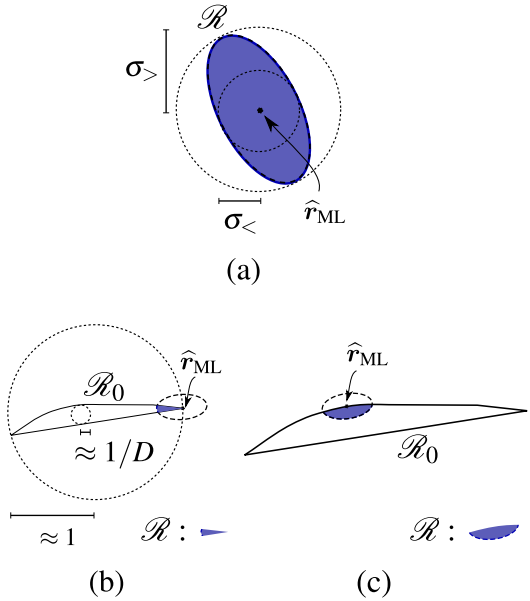}
	\caption{\label{fig:complexity}(Color Online) Schematic diagrams for the geometrical relationship between the CR $\mathcal{R}$ and the quantum state space $\mathcal{R}_0$. The situation for (a) Case~A is completely known and so complexity estimation for hit-and-run is a simple matter. To acquire conservative complexity estimates for Case~B, two special types of such CR may exist: either the CR (b) lies on an extremely sharp corner of $\mathcal{R}_0$ in at least one of its dimension~(Type I) in whichever orientation, or (c)~on one of its edges that is almost flat (Type II) in all its dimensions, with the longest $\mathcal{E}'_\lambda$-axis oriented along the flat surface.}
\end{figure}

To further speed up the algorithm for Case~B, one can assign $\mathcal{B}$ to be the hyperellipsoidal cap composed by a hyperplane that is tangent to the isoGaussian level curve of $\mathcal{E}'_\lambda$ at $\ML{\rvec{r}}$ and the part of $\mathcal{E}'_\lambda$ below it~(refer to Sec.~\ref{subsec:hecap}). Numerical experience shows that this speed up is negligible in the presence of the endpoint adaptation mechanism of accelerated hit-and-run. 

\suggtwo{We end this introduction of hit-and-run by noting that a hyperellipsoidal $\mathcal{B}$ is constructed based on the large-$N$ limit, where the boundary $\partial\mathcal{R}_\lambda\cap\mathrm{int}\{\mathcal{R}_0\}$ of the physical region is well approximated by this hyperellipsoid. The highly skeptical may insist that, perhaps, for a finite $N$, even if $N\gg1$, there might still be cases where a part of this boundary protrudes $\mathcal{B}$. To be on the safer side, one may choose a hyperellipsoidal $\mathcal{B}$ of a reasonably larger size (say doubled) than the one given by the central limit theorem. This will almost surely contain the physical error region with a much smaller failure probability. The pertinent question is: ``Can we verify that $\mathcal{B}$ contains $\mathcal{R}$ with arbitrary precision?'' The answer unfortunately is negative both in theory and practice. This is because a positive answer would entail a complete knowledge about $\partial\mathcal{R}$, obtaining which is either computationally not feasible in general, or an NP-hard problem in some context~\cite{Suess:2017np}.}

\subsection{Numerical complexity estimations}
\label{subsec:numcmpl}

After suppressing dependences on logarithmic factors and error parameters, it was argued that the number of hit-and-run steps needed to gather enough sample points and form an ensemble described by $p(\rvec{r})$ in hit-and-run is $O\!\left(d^2 R^2_\text{out}/R^2_\text{in}\right)=O\!\left(D^4 R^2_\text{out}/R^2_\text{in}\right)$~\cite{Cousins:2016va,Lovasz:1999aa} in the limit $D\gg2$, where $R_\text{out}$ is the radius of the smallest outer sphere that contains $\mathcal{R}_\lambda$ and $R_\text{in}$ is that of the largest inner sphere that can be inscribed in $\mathcal{R}_\lambda$. Together with the floating-point-operations complexity $O(D^3)$ in a typical Cholesky decomposition algorithm~\cite{Higham:2009aa}, we have an estimate for the complexity $\cmpl=O(D^7R^2_\text{out}/R^2_\text{in})$ for the entire hit-and-run scheme.

The treatment of Case~A is straightforward as we have the complete information about $\mathcal{R}_\lambda\approx\mathcal{E}_\lambda$ in the large-$N$ limit. If we denote $\sigma_>$ and $\sigma_<$ to respectively be the largest and smallest eigenvalue of $\widetilde{\FML}^{-1/2}$, then the corresponding outer and inner radii are $R_\text{out}=\sigma_>$ and $R_\text{in}=\sigma_<$ [see Fig.~\ref{fig:complexity}(a)], so that $\cmpl_\text{A}=O\!\left(D^7\mathrm{cond}\!\left(\FML^{-1}\right)\right)$ \sugg{involves the conditional number $\mathrm{cond}\!\left(\FML^{-1}\right)=\sigma_>^2/\sigma_<^2$}.

The analysis for Case~B requires extra care given the complicated state-space boundary $\partial\mathcal{R}_0$. While complete and precise details of $\mathcal{R}_0$ are absent so far, from \cite{Bengtsson:2013gm}, we know that in the Euclidean space, the largest inner sphere inscribable in $\mathcal{R}_0$ has a radius that approaches $1/D$ for $D\gg2$, and that the smallest outer sphere that contains $\mathcal{R}_0$ has a radius going to 1 in the same dimension limit. The overall shape of $\mathcal{R}_0$ is therefore a ``squashed'' convex body for large $D$, such that \emph{at least one} of its dimensions drops appreciably to zero. To estimate the complexity for Case~B, we consider CRs of two tractable types: a Type~I CR is located at an extremely sharp corner of $\mathcal{R}_0$ that is made from at least one of its rapidly shrinking dimensions, as shown in Fig.~\ref{fig:complexity}(b), whereas a Type~II CR is situated at an extremely flat boundary of $\mathcal{R}_0$ where all of its dimensions remain approximately constant within the CR as in Fig.~\ref{fig:complexity}(c). For a conservative estimate of $\cmpl$, we consider an $\mathcal{R}$ such that the longest axis of $\mathcal{E}'_\lambda$ is aligned with the flat surface. All other types of Case-B CRs may be viewed as intermediate situations of these two and have no analytical complexity estimates known to us. The data-copy number $N\gg1$ is assumed to be sufficiently large such that $\gML\approx\rvec{0}$ and $\rvec{r}_\text{c}\approx\ML{\rvec{r}}$.

\begin{figure}[t]
	\center
	\includegraphics[width=0.82\columnwidth]{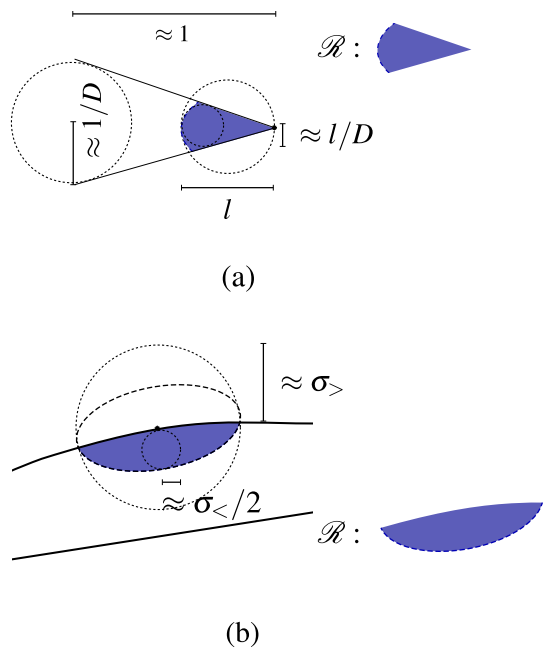}
	\caption{\label{fig:typeI_II}(Color Online) Schematic diagrams for (a)~Type~I and (b)~Type~II Bayesian regions. Type I regions have complexities that are strongly influenced by the cornered geometry (greatly exaggerated for visual aid), whereas Type II regions have complexities that strongly depends on the eigenvalue aspect ratios of $\FML$. All other intermediate CR types give rise to complexities affected by the geometries of both $\partial\mathcal{R}_0$ and $\FML$.}
\end{figure}

To estimate $\cmpl$ for a Type~I CR, we assume that the corner is extremely sharp in one particular dimension such that the curvature of $\partial\mathcal{R}_0$ extending out from $\ML{\rvec{r}}$ is almost flat. Then following Fig.~\ref{fig:typeI_II}(a), the concept of similar figures give $R_\text{out}/R_\text{in}\approx D$, which is independent of $\FML$ for extremely sharp corners, and $\cmpl_{\text{B,I}}=O\!\left(D^9\right)$. The complexity for Type~II CRs may be estimated with the help of Fig.~\ref{fig:typeI_II}(b), where  $R_\text{out}/R_\text{in}\approx 2\,\mathrm{cond}\!\left(\FML^{-1/2}\right)$ is now independent of $\partial\mathcal{R}_0$ due to its extremely mild edge features, leading us to $\cmpl_{\text{B,II}}=O\!\left(D^7\mathrm{cond}\!\left(\FML^{-1}\right)\right)=\cmpl_{\text{A}}$.

\subsection{Other numerical methods}
\sugg{
Other numerical methods apart from hit-and-run may also be used to perform in-region sampling, each of which has its own merits and shortcomings~\cite{Shang:2015mc,DelMoral:2006aa,Metropolis:1953aa,Hastings:1970aa}. Classical rejection and importance sampling methods are two straightforward ways to acquire samples distributed according to some desired prior distribution. For large $D$ and $N$, these methods rapidly become infeasible due to the decreasing ratio of the CR volume to the full Hermitian sampling volume that includes many more unphysical operators that are not quantum states. To cope with this low-yield difficulty, another Markov-chain method known as the Metropolis-Hastings MC scheme may also be used to do in-region sampling. Such a scheme also suffers from high correlations that are generally dependent on the starting point of a Markov-chain iteration. Hamiltonian MC methods are yet another promising class of algorithms that permit larger sample-point hopping that gives a sample with weak correlations. The scalability of such methods are, however, still work in progress.}

\section{Distance-induced region capacity}
\label{sec:newsize}

\subsection{The operational definition}

The \sugg{theory of in-region sampling seamlessly} paves the way to other creative ways of defining the \sugg{capacity of a region}. Doing so permits us to talk about \sugg{``how big'' a CR is without referring} to $\mathcal{R}_0$ entirely. To begin, one could measure the region capacity in terms of the average distance between any two points inside $\mathcal{R}$\sugg{, which is a separate idea from the prior content}. Intuitively, the smaller this average distance, the smaller the region and \emph{vice versa}. Using this simple prescription, we propose the region-average quantity
\begin{equation}
S_{\mathcal{D},\lambda}\equiv\overline{\mathcal{D}(\rvec{r}',\ML{\rvec{r}})}^{\mathcal{R}_\lambda}=\dfrac{\displaystyle\int_{\mathcal{R}_\lambda}(\D\,\rvec{r}')\,\mathcal{D}(\rvec{r}',\ML{\rvec{r}})}{\displaystyle\int_{\mathcal{R}_\lambda}(\D\,\rvec{r}')}
\label{eq:Slbd}
\end{equation}
to measure the \sugg{capacity} of $\mathcal{R}_\lambda$, where $\mathcal{D}(\rvec{r}',\ML{\rvec{r}})$ is some pre-chosen distance metric. Notice that the ML estimator $\ML{\rvec{r}}$ is selected to be the reference point from which distances are measured without loss of generality. 

To concretize all results, we shall look at three distance measures for states that enjoy a good reputation in quantum-information studies. We first mention the Hilbert-Schmidt (HS) distance
\begin{equation}
\mathcal{D}_\textsc{hs}=\tr{(\rho'-\ML{\rho})^2}=(\rvec{r}'-\ML{\rvec{r}})^2\,,
\label{eq:hs_dist}
\end{equation}
which is equivalent to the squared $l_2$-norm of $\rvec{r}'-\ML{\rvec{r}}$. Closely related to the HS distance is the trace-class distance
\begin{equation}
\mathcal{D}_\text{tr}=\tr{\sqrt{(\rho'-\ML{\rho})^2}}
\label{eq:tr_dist}
\end{equation}
defined by the operator absolute value $|A|=\sqrt{A^\dagger A}$. To introduce the third measure, we start by quoting the expression of quantum fidelity~\cite{Uhlmann:1976aa}
\begin{equation}
\mathcal{F}=\tr{\sqrt{\sqrt{\ML{\rho}}\rho'\sqrt{\ML{\rho}}}}^2
\label{eq:fid}
\end{equation}
between $\rho'$ and $\ML{\rho}$, to which we can define the Bures distance~\cite{Bures:1969aa,Braunstein:1994aa} $\mathcal{D}_\textsc{b}=2(1-\sqrt{\mathcal{F}})$. In the limit of large $N$, where the fidelity $\mathcal{F}\approx1-\epsilon$ differs from 1 by a small amount, $\mathcal{D}_\textsc{b}$ is also approximately the infidelity $1-\mathcal{F}$.

\begin{figure}[t]
	\center
	\includegraphics[width=1\columnwidth]{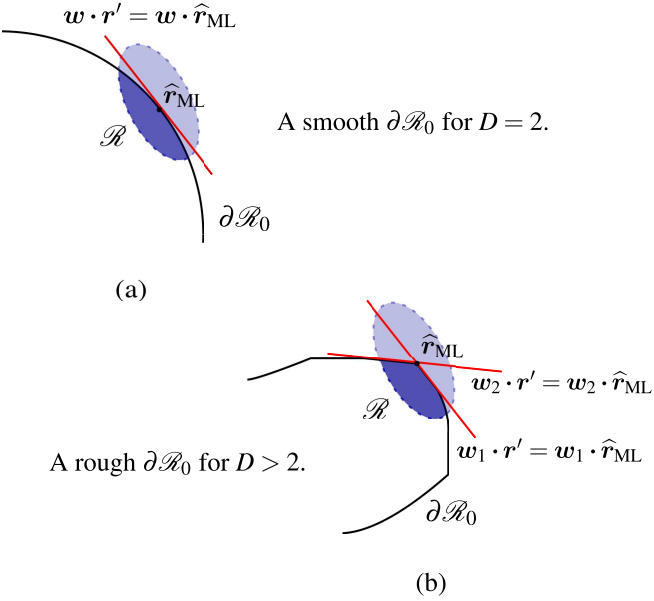}
	\caption{\label{fig:bd}(Color Online) The different types of state-space boundaries $\partial\mathcal{R}_0$~\sugg{\cite{PRL}}. Excluding the exceptional single-qubit system that exhibits a smooth spherical surface~(a), all higher-dimensional systems result in $\partial\mathcal{R}_0$ that is not smooth, with corners and edges. In the large-$N$ limit, an ML estimator at the corner, for instance, may be well-approximated by a collection of hyperplanes relative to $\mathcal{R}$ since every point in $\partial\mathcal{R}_0$ is a well-defined quantum state.}
\end{figure}

\subsection{Monotonic behavior of $S_{\mathcal{D},\lambda}$ for $N\gg1$}

Here, we show that, at least for sufficiently large $N$, $S_{\mathcal{D},\lambda}$, defined by any of these three distance measures, behave correctly as a \sugg{capacity} function in the sense that $S_{\mathcal{D},\lambda}$ should not increase as $\lambda$ increases. We first look at the more complicated Case~B, and argue that since the $\mathcal{R}_0$ generally has only corners and edges with no other mathematically pathological features, a set of hyperplanes can then be used to model any particular boundary feature on which $\ML{\rvec{r}}$ resides (see Fig.~\ref{fig:bd}). This results in the asymptotic form
\begin{equation}
S_{\mathcal{D},\lambda}\rightarrow\dfrac{\displaystyle\int(\D\,\rvec{r}'')\,\mathcal{D}\,\eta(1-\TP{\rvec{r}''}\FML\rvec{r}''/(-2\log\lambda))\prod_j\eta(\rvec{w}^\textsc{t}_j\rvec{r}'')}{\displaystyle\int(\D\,\rvec{r}'')\eta(1-\TP{\rvec{r}''}\FML\rvec{r}''/(-2\log\lambda))\prod_j\eta(\rvec{w}^\textsc{t}_j\rvec{r}'')}
\label{eq:Sbd}
\end{equation}
after the substitution $\rvec{r}''=\rvec{r}'-\ML{\rvec{r}}$.

At this stage, we shall consider the asymptotic expressions of the distance measures. The HS distance $\mathcal{D}_\textsc{hs}$ takes on the simplest (quadratic) form out of all three, which very straightforwardly gives the asymptotic dependence $S_{\textsc{hs},\lambda}\rightarrow-\log\lambda$ provided the sufficient condition $(\D\,\alpha\,\rvec{r}')=g(\alpha)\,(\D\,\rvec{r}')$, which includes the \emph{uniform primitive prior} $(\D\,\rvec{r}')=(\D\,\rvec{r}')_\text{unif}\equiv\prod_j\D\,r_j$. It is not difficult to see that the same $\lambda$ dependence applies to Case~A by taking $\rvec{w}_j=\rvec{0}$, so that $S_{\textsc{hs},\lambda}$ is monotonically decreasing with increasing $\lambda$. Next, according to Appendix~\ref{app:hsandtr}, in the limit of large $D$, $S_\text{tr}\sim\sqrt{S_\textsc{hs}}$, which is also clearly monotonic as well owing to $S_\textsc{hs}$'s monotonicity. For $S_\textsc{b}$ and $S_\mathcal{F}$, one can perform a Taylor expansion on them about $\ML{\rvec{r}}$ (see Sec.~\ref{sec:formulas}) and realizes that both functions asymptotically depend on the dyadic $(\rvec{r}'-\ML{\rvec{r}})(\rvec{r}'-\ML{\rvec{r}})^\textsc{t}$, so that both region-average distances are also asymptotically decreasing with $\lambda$.

\section{Approximation formulas for $S_{\mathcal{D},\lambda}$ and $u_\lambda$}
\label{sec:formulas}

The prior content $S_\lambda$ discussed alongside $C_\lambda$ in Secs.~\ref{sec:std_th}--\ref{sec:numcomp} quantifies the size of $\mathcal{R}_\lambda$ relative to $\mathcal{R}_0$. In our earlier article~\cite{Teo:2018aa}, analytical approximation formulas for $S_\lambda$ were proposed in the large-$N$ limit, all of which are scaled with the volume $V_{\mathcal{R}_0}$ of $\mathcal{R}_0$. As is also shown later in the section, this volume dependence is associated with the extension of every $\mathcal{R}_0$ integral
\begin{equation}
\int_{\mathcal{R}_0}(\D\,\rvec{r}')\,\eta(L-\lambda\,L_\text{max})\cdots\rightarrow\dfrac{1}{V_{\mathcal{R}_0}}\int_{\text{all space}}\prod_j\D\,r'_j\cdots
\end{equation}
to the entire $\rvec{r}'$ space ascribed with the uniform primitive prior, which is a reasonable step to obtain analytical results under the central limit theorem since $L$ is narrow enough to reside within the confines of $\mathcal{R}_\lambda$ under this limit. Therefore, the valid usage of these theoretical expressions hinges on the availability of $V_{\mathcal{R}_0}$. In quantum-state tomography where we have no complete theoretical information about $\mathcal{R}_0$,  $V_{\mathcal{R}_0}$ is known only for certain priors and state parametrizations ~\cite{Zyczkowski:2003hs,Andai:2006aa,Bengtsson:2006gm,Bengtsson:2013gm}.

On the other hand, it is obvious that $V_{\mathcal{R}_0}$ is canceled out for any region-average quantity after such integral extensions. This allows one to derive operational asymptotic formulas for averages like $S_{\mathcal{D},\lambda}$ and $u_\lambda$ regardless of $\mathcal{R}_0$ in whichever parametrization. As a calculable standard in this section, we continue to derive expressions in terms of the uniform primitive prior and $\rvec{r}$, although the subsequent instructions may also work for other manageable priors with which $S_{\mathcal{D},\lambda}$ behaves as a proper region-capacity function. We first address the different $\mathcal{D}$ measures in the large-$N$ limit. 

\subsection{The various $\mathcal{D}$ measures}
\label{subsec:Dmeas}

\subsubsection{Hilbert-Schmidt and trace-class measures}

The HS distance measure $\mathcal{D}_
\textsc{hs}(\rvec{r}',\ML{\rvec{r}})$ takes the very simple quadratic form in \eqref{eq:hs_dist} under any circumstance, whereas the trace-class distance $\mathcal{D}_\text{tr}$ has no easy functional form in terms of $\rvec{r}'$ for $D>2$. Nevertheless in the limits $N\gg1$ \emph{and} $D\gg2$, based on the principles of random matrix theory detailed in Appendix~\ref{app:hsandtr}, it is deduced that the asymptotic expression
\begin{equation}
S_\text{tr}\approx\dfrac{8\sqrt{D\,S_\textsc{hs}}}{3\pi}
\label{eq:tr2hs}
\end{equation}
relating the final $\mathcal{R}$-averages $S_\textsc{hs}$ and $S_\text{tr}$ is approximately valid for both Case~A and B, which incidentally takes the same form found in \cite{Zhu:2011sp} that was calculated for statistical-fluctuation studies.

\subsubsection{Bures measure}

The Bures distance measure $\mathcal{D}_\textsc{b}$ also has no tractable functional form in $\rvec{r}'$ for general $D$. To find the asymptotic link with $\rvec{r}'$ this time, it is technically more convenient to inspect the behavior of $\mathcal{F}$ around $\ML{\rho}\leftrightarrow\ML{\rvec{r}}$ as $N\gg1$. 

A Taylor expansion about $\ML{\rho}$ as guided in Appendix~\ref{app:fid}, we have
\begin{equation}
\mathcal{F}_\text{A}\approx 1-\dfrac{1}{2}\TP{(\rvec{r}'-\ML{\rvec{r}})}\,\dyadic{Q}_D\,(\rvec{r}'-\ML{\rvec{r}})
\end{equation}
for Case~A and
\begin{align}
\mathcal{F}_\textsc{b}\approx&\, 1+\TP{(\rvec{r}'-\ML{\rvec{r}})}\tr{P_r\,\rvec{\Omega}}\nonumber\\
&\!\!\!\!+\dfrac{1}{2}\TP{(\rvec{r}'-\ML{\rvec{r}})}\,\left(\dfrac{1}{2}\tr{P_r\,\rvec{\Omega}}\tr{P_r\,\TP{\rvec{\Omega}}}-\dyadic{Q}_r\right)\,(\rvec{r}'-\ML{\rvec{r}})
\end{align}
for Case~B, where $P_r$ is the projector onto the support of $\ML{\rho}$ \sugg{having the rank-deficient spectral decomposition $\ML{\rho}=\sum^r_{j=1}\ket{\lambda_j}\lambda_j\bra{\lambda_j}$}, and 
\begin{equation}
\dyadic{Q}_r=\sum^r_{j=1}\sum^r_{k=1}\dfrac{\opinner{\lambda_j}{\rvec{\Omega}}{\lambda_k}\opinner{\lambda_k}{\TP{\rvec{\Omega}}}{\lambda_j}}{\lambda_j+\lambda_k}\,.
\end{equation}

\subsection{Case~A: hyperellipsoidal theory}

The presentation in Sec.~\ref{subsec:Dmeas} reduces the necessary ingredients for \sugg{large-$N$ (or $D$)} analytical estimations of $S_{\mathcal{D},\lambda}$ to just the scalar $\int_{\mathcal{R}_\lambda}(\D\,\rvec{r}')$, column $\int_{\mathcal{R}_\lambda}(\D\,\rvec{r}')\,\rvec{\Delta}'_\textsc{ml}$ and dyadic $\int_{\mathcal{R}_\lambda}(\D\,\rvec{r}')\,\rvec{\Delta}'_\textsc{ml}\,\TP{\rvec{\Delta}'}_\textsc{ml}$, where $\rvec{\Delta}'_\textsc{ml}=\rvec{r}'-\ML{\rvec{r}}$.

When $\mathcal{R}_\lambda\approx\mathcal{E}_\lambda$, these three integrals takes on simple analytical forms. We start with
\begin{align}
\int_{\mathcal{R}_\lambda}(\D\,\rvec{r}')=&\,\int_{\mathcal{R}_0}(\D\,\rvec{r}')\,\eta(1-\TP{\rvec{\Delta}'}_\textsc{ml}\FML\rvec{\Delta}'_\textsc{ml}/(-2\,\log\lambda))
\end{align}
and transform $\rvec{r}'\rightarrow\rvec{r}''=\dyadic{D}^{1/2}\,\TP{\dyadic{O}}\rvec{\Delta}'_\textsc{ml}$ to the translated diagonal coordinate variables of $\FML/(-2\,\log\lambda)=\dyadic{O}\,\dyadic{D}\,\TP{\dyadic{O}}$, so that in the large-$N$ limit and uniform primitive prior, we may relax the boundary of $\mathcal{R}_0$ and write
\begin{align}
\int_{\mathcal{R}_\lambda}(\D\,\rvec{r}')\rightarrow&\,\dfrac{\DET{\dyadic{D}^{-1/2}}}{V_{\mathcal{R}_0}}\int\,(\D\,\rvec{r}'')_\text{unif}\,\eta(1-\rvec{r}''^2)\nonumber\\
=&\,\dfrac{V_d}{V_{\mathcal{R}_0}}\,(-2\log\lambda)^{d/2}\,\DET{\FML}^{-1/2}\,,
\end{align}
which is a function of the volume $V_d=\pi^{d/2}/(d/2)!$ of the $d$-dimensional unit hyperball, the inverse of $\FML$ that characterizes $\mathcal{E}_\lambda$ together with the logarithm of $\lambda$.

In this case, the integral column is zero since the integrand after variable transformation becomes odd in $\rvec{r}''$, and we are thus left with
\begin{align}
\int_{\mathcal{R}_\lambda}(\D\,\rvec{r}')\,\rvec{\Delta}'_\textsc{ml}\,\TP{\rvec{\Delta}'}_\textsc{ml}\rightarrow&\,\dfrac{\DET{\dyadic{D}^{-1/2}}}{V_{\mathcal{R}_0}}\,\dyadic{O}\,\dyadic{D}^{-1/2}\,\dyadic{I}\,\dyadic{D}^{-1/2}\,\TP{\dyadic{O}}\,,
\end{align}
and
\begin{align}
\dyadic{I}=&\,\int\,(\D\,\rvec{r}'')_\text{unif}\,\eta(1-\rvec{r}''^2)\,\rvec{r}''\,\TP{\rvec{r}''}\nonumber\\
=&\,\int_\text{unit sphere}\,(\D\,\rvec{r}'')\,\rvec{r}''\,\TP{\rvec{r}''}\nonumber\\
=&\,\int^1_0\,\D\,r''\,r''^{\,d+1}\,\int\,(\D\,\{\text{solid angle}\})\,\rvec{e}''\,\TP{\rvec{e}''}=\dfrac{V_d}{d+2}\dyadic{1}\,,
\end{align}
where the last equality is explained by the orthogonally invariant of the $(d-1)$-dimensional solid-angle measure over the unit columns $\rvec{e}$, and so
\begin{equation}
\int_{\mathcal{R}_\lambda}(\D\,\rvec{r}')\,\rvec{\Delta}'_\textsc{ml}\,\TP{\rvec{\Delta}'}_\textsc{ml}\rightarrow\dfrac{V_d}{V_{\mathcal{R}_0}}\,(-2\log\lambda)^{d/2+1}\DET{\FML}^{-1/2}\,\invFML\,.
\label{eq:dyadic_A}
\end{equation}

With all these components, the relevant asymptotic formulas concerning all three distance measures
\begin{alignat}{3}
&\,S^\text{ (A)}_{\textsc{hs},\lambda}&&\approx&&\,\,\Tr{\invFML}\dfrac{-\log\lambda}{d/2+1}\,,\nonumber\\
&\,S^\text{ (A)}_{\text{tr},\lambda}&&:&&\,\,\text{as in \eqref{eq:tr2hs}}\,,\nonumber\\
&\,S^\text{ (A)}_{\textsc{b},\lambda}&&\approx&&\,\,\Tr{\invFML\,\dyadic{Q}_D}\dfrac{-\log\lambda}{d+2}\,.
\label{eq:caseA_dist}
\end{alignat}
Here $\mathrm{Tr}$ now addresses the dyadic character, as opposed to $\mathrm{tr}$, and we witness the manifestation of logarithmic divergences from both the relaxation of $\partial\mathcal{R}_0$ and Gaussian approximation of $L$.

Next, to analytically calculate \sugg{$u_\lambda$ using $f(L)=\log L$} with which $C_\lambda$ can be found, we note that due to the Gaussian form of $L$,\sugg{
\begin{equation}
u_\lambda=-\log \lambda-\dfrac{\displaystyle\int_{\mathcal{R}_\lambda}(\D\,\rvec{r}')\,\TP{\rvec{\Delta}'}_\textsc{ml}\,\FML\,\rvec{\Delta}'_\textsc{ml}}{\displaystyle 2\int_{\mathcal{R}_\lambda}(\D\,\rvec{r}')}
\label{eq:caseA_u}
\end{equation}
}is a dyadic trace function of $\int_{\mathcal{R}_\lambda}(\D\,\rvec{r}')\,\rvec{\Delta}'_\textsc{ml}\,\TP{\rvec{\Delta}'}_\textsc{ml}$, so that we may use the right-hand side of \eqref{eq:dyadic_A} and put down\sugg{
\begin{equation}
u_{\text{A},\lambda}=-\dfrac{2}{d+2}\log\lambda
\end{equation}
}after some basic trace and logarithmic manipulations. It is clear that $d/(d+2)\leq u_{\text{A},\lambda}\leq1$ is bounded. 

\subsection{Case~B: hyperellipsoidal-cap theory}
\label{subsec:hecap}

\begin{figure}[t]
	\center
	\includegraphics[width=1\columnwidth]{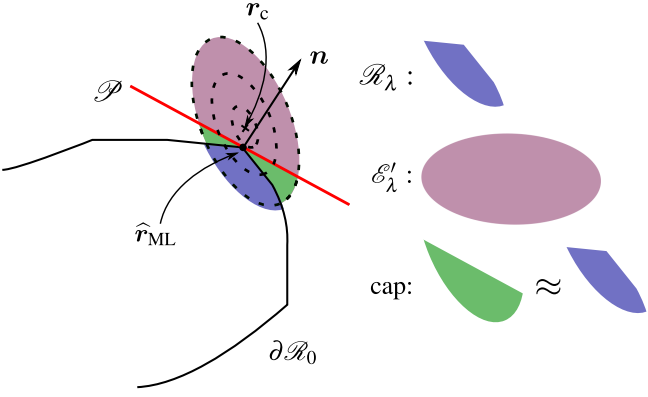}
	\caption{\label{fig:cap}(Color Online) Modeling the boundary $\partial\mathcal{R}_\lambda\cap\partial\mathcal{R}_0$~\sugg{\cite{PRL}}: A hyperplane $\mathcal{P}$ (red solid line) is introduced in a manner that its normal $\rvec{n}$ is orthogonal to the level curve at $\ML{\rvec{r}}$ to form a cap that approximates $\mathcal{R}_\lambda$.}
\end{figure}

In Case~B, although the geometry of $\mathcal{R}_\lambda\approx\mathcal{E}'_\lambda\cap\mathcal{R}_0$ is now much trickier to deal with, the central limit theorem proposed in Sec.~\ref{sec:bigdata} allows us to approximate $\mathcal{R}_\lambda$ by a regular analytical region.

As shown in Fig.~\ref{fig:cap}, one can introduce a hyperplane $\mathcal{P}$, described by $\rvec{n}\bm{\cdot}(\rvec{r}'-\ML{\rvec{r}})=0$ ($\rvec{n}\propto\gML$) that is tangent to the level curve of the Gaussian function in \eqref{eq:caseB_clt} at $\ML{\rvec{r}}$. The hyperspherical cap formed by $\mathcal{P}$ and $\mathcal{E}'_\lambda$ hence asymptotically contains $\mathcal{R}_\lambda$, where we have essentially modeled the highly nontrivial $\partial\mathcal{R}_\lambda\cap\partial\mathcal{R}_0$ as $\mathcal{P}$. This model implies the estimated assignment
\begin{align}
&\,\int_{\mathcal{R}_\lambda}(\D\,\rvec{r}')\cdots\nonumber\\
=&\,\dfrac{1}{V_{\mathcal{R}_0}}\int(\D\,\rvec{r}')_\text{unif}\,\eta(1-\TP{(\rvec{r}'-\rvec{r}_\text{c})}\,\FML\,(\rvec{r}'-\rvec{r}_\text{c})/(-2\,\log\lambda'))\nonumber\\
&\,\qquad\qquad\times\eta(\rvec{n}\bm{\cdot}(\ML{\rvec{r}}-\rvec{r}'))\cdots\,.
\end{align}
The change of variable $\rvec{r}'\rightarrow\rvec{r}''=\dyadic{D}'^{1/2}\,\TP{\dyadic{O}'}\,(\rvec{r}'-\rvec{r}_\text{c})$ with respect to the diagonal coordinates of $\FML/(-2\log\lambda')=\dyadic{O}'\,\dyadic{D}'\,\TP{\dyadic{O}'}$ leads to
\begin{equation}
\int_{\mathcal{R}_\lambda}(\D\,\rvec{r}')\,q_\lambda(\rvec{r}')\approx\dfrac{\DET{\dyadic{D}'^{-1/2}}}{V_{\mathcal{R}_0}}\int(\D\,\rvec{r}'')_\text{cap}\,q_\lambda(\dyadic{O}'\,\dyadic{D}'^{-1/2}\,\rvec{r}'')
\end{equation}
for any function $q$, which is parametrized by the cap element $(\D\,\rvec{r}'')_\text{cap}=(\D\,\rvec{r}'')_\text{unif}\,\eta(1-\rvec{r}''^2)\,\eta(a-\TP{\rvec{b}}\,\rvec{r}'')$, $a=\TP{\rvec{n}}(\ML{\rvec{r}}-\rvec{r}_\text{c})$ and $\rvec{b}=\dyadic{D}'^{-1/2}\,\TP{\dyadic{O}'}\,\rvec{n}$. One can check that
\begin{align}
l\equiv\dfrac{a}{|\rvec{b}|}=&\,\dfrac{\gML\bm{\cdot}(\ML{\rvec{r}}-\rvec{r}_\text{c})}{|\dyadic{D}'^{-1/2}\,\TP{\dyadic{O}'}\,\gML|}\nonumber\\
=&\,\sqrt{\dfrac{\TP{(\ML{\rvec{r}}-\rvec{r}_\text{c})}\,\FML\,(\ML{\rvec{r}}-\rvec{r}_\text{c})}{(-2\,\log\lambda')}}\leq 1\,.
\end{align}

In other words, we have
\begin{equation}
\overline{q_\lambda(\rvec{r}')}^{\mathcal{R}_\lambda}\approx\dfrac{\displaystyle\int(\D\,\rvec{r}'')_\text{cap}\,q_\lambda(\dyadic{O}'\,\dyadic{D}'^{-1/2}\,\rvec{r}'')}{\displaystyle\int(\D\,\rvec{r}'')_\text{cap}}\,,
\end{equation}
and that for any $q_\lambda$ belonging to either one of the three distance measures or \sugg{$\log L-\log(\lambda L_\text{max})$}, as reasoned in Sec.~\ref{subsec:hecap}, the building blocks of $\overline{q_\lambda(\rvec{r}')}^{\mathcal{R}_\lambda}$ are only $\int(\D\,\rvec{r}'')_\text{cap}$,  $\int(\D\,\rvec{r}'')_\text{cap}\,\rvec{r}''$ and $\int(\D\,\rvec{r}'')_\text{cap}\,\rvec{r}''\,\TP{\rvec{r}''}\vphantom{{M^M}^M}$. These integrations are all carried out in Appendix~\ref{app:cap}.

\begin{figure}[t]
	\center
	\includegraphics[width=1\columnwidth]{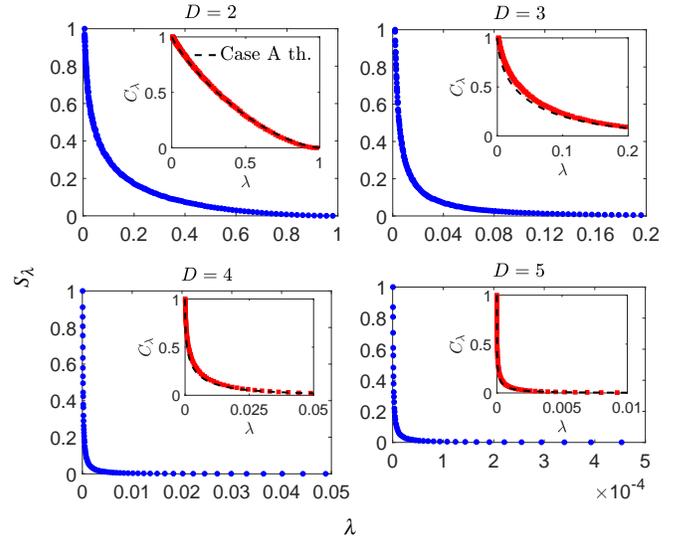}
	\caption{\label{fig:uSCA}(Color Online) \sugg{Graphs of size $S_\lambda$ and credibility $C_\lambda$ for Case~A with $2\leq D\leq5$. $M$, the number of POM outcomes, is set to $D^3$ and the POM is chosen to be a random square-root measurement as a simulation example for each $D$. Here $N/M=500$. The dashed curve in every inset is computed with the Case~A large-$N$ formula for $C_\lambda$ in Eq.~(7) of \cite{Teo:2018aa}.} A randomly chosen rank-$D$ true state $\rho$ is used in each panel. \sugg{A total of 20000 points are collected during in-region sampling of $u_\lambda$ per $\lambda$.}}
\end{figure}

In combining all results gathered from Appendices~\ref{app:fid} and \ref{app:cap}, we denote $\mathcal{N}_{d,l,x}=V_d\,\mathrm{I}_{(1-l)/2}((d+x)/2,(d+x)/2)$, which depends on the incomplete Euler's beta function $\mathrm{I}_\cdot(\cdot,\cdot)$, and organize two new auxiliary quantities
\begin{align}
\rvec{m}=&\,\left[-\dfrac{V_{d-1}}{l(d+1)}\left(1-l^2\right)^{(d+1)/2}+\mathcal{N}_{d,l,1}\right]\FML^{-1}\gML\,,\nonumber\\
\dyadic{M}=&\,\dfrac{-\log\lambda'}{d+2}\,\mathcal{N}_{d,l,3}\,\FML^{-1}+\dfrac{1}{2}\,\rvec{m}\,\gML^\textsc{t}\FML^{-1}\,.
\label{eq:auxop}
\end{align}
This helps to clean the respective formulas
\begin{alignat}{3}
&\,S^\text{ (B)}_{\textsc{hs},\lambda}&&\approx&&\,\,\dfrac{\Tr{2\,\dyadic{M}}}{\mathcal{N}_{d,l,1}}\,,\nonumber\\
&\,S^\text{ (B)}_{\text{tr},\lambda}&&:&&\,\,\text{as in \eqref{eq:tr2hs}}\,,\nonumber\\
&\,S^\text{ (B)}_{\textsc{b},\lambda}&&\approx&&\,\,\dfrac{\tr{P_r\,\TP{\rvec{\Omega}}\rvec{m}}+\Tr{\dyadic{M}\,\dyadic{Q}_r}}{\mathcal{N}_{d,l,1}}\approx\dfrac{\tr{P_r\,\TP{\rvec{\Omega}}\rvec{m}}}{\mathcal{N}_{d,l,1}}\,,
\label{eq:caseB_dist}
\end{alignat}
for the distance-induced capacity functions and\sugg{
\begin{align}
u_{\text{B},\lambda}=&\,[-\log \lambda'+\Tr{\gML\rvec{m}^\textsc{t}-\FML\,\dyadic{M}}/\mathcal{N}_{d,l,1}]\nonumber\\
&\,\times\log(\lambda\,L_\text{max})/\log(\lambda'\,L_\text{max})\,.
\label{eq:caseB_u}
\end{align}
}We caution the Reader once more regarding the actions of $\mathrm{tr}$ and $\mathrm{Tr}$ at the right-hand side of $S^\text{ (B)}_{\textsc{b},\lambda}$ in \eqref{eq:caseB_dist}.

For consistency, we end this section by noting that Eqs.~\eqref{eq:caseB_dist} and \eqref{eq:caseB_u} cover Eqs.~\eqref{eq:caseA_dist} and \eqref{eq:caseA_u} because Case~A implies that $\lambda'=\lambda$ ($\gML=\rvec{0}=\rvec{m}$), such that $l=0$ then gives $\mathcal{N}_{d,0,x}=V_d$ and $\dyadic{M}=(-\log\lambda)\,\FML^{-1}/(d+2)$. 

\section{Results and discussions}
\label{sec:results}

\begin{figure}[t]
	\center
	\includegraphics[width=1\columnwidth]{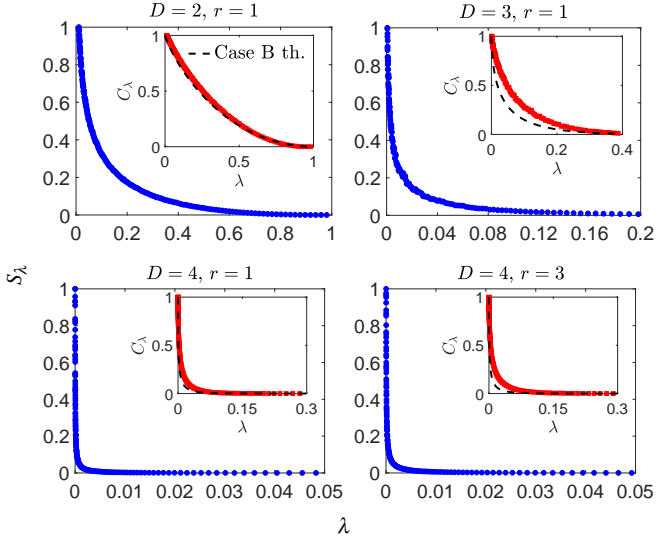}
	\caption{\label{fig:uSCB}(Color Online) \sugg{Graphs of computed $S_\lambda$ and $C_\lambda$ Case~B against $\lambda$, with $2\leq D\leq4$, which have, otherwise, the same specifications as Fig.~\ref{fig:uSCA}. The rank $r$ of $\ML{\rho}$, which characterizes a pure state, for each panel is explicitly stated. The dashed curves in this figure are generated from the Case~B formulas in Eq.~(14) of \cite{Teo:2018aa}.}}  
\end{figure}

\subsection{Region reconstruction}

We first present, \sugg{under the uniform primitive prior $p(\rvec{r})\propto1$}, the computation results of $S_\lambda$ and $C_\lambda$ from $u_\lambda$ in Figs.~\ref{fig:uSCA} and \ref{fig:uSCB} for quantum systems of various dimensions $D$. To be more technically precise about our use of Euler's method described in Sec.~\ref{sec:newtheory}, we first solve \eqref{eq:ode2} for $y_\lambda$ by iterating \eqref{eq:euler} starting with a numerically small $\lambda$ value to $\lambda\approx1$ using the \sugg{function $f(\lambda)=\log L$}. 

The behavior of $S_\lambda$ shows the expected decreasing trend not only in $\lambda$, but also in overall magnitude as $D$ increases. \sugg{This indicates that the (log-)likelihood is turning into a delta-function peak.} For larger $D$ or $N$, the computational accuracy of $S_\lambda$ and $C_\lambda$ using Euler's numerical method may be maintained by exploring many more $\lambda$ values near zero, such as \emph{via} logarithmic scaling of $\lambda$, as all curves possess sharp gradient changes in this $\lambda$ range. 

\begin{figure}[t]
	\center
	\includegraphics[width=1\columnwidth]{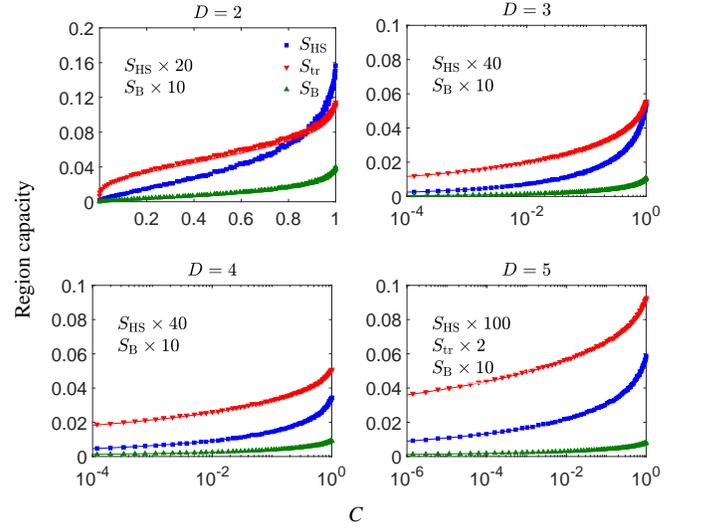}
	\caption{\label{fig:SCA}(Color Online) Graphs of the distance-induced \sugg{region-capacity function} against the credibility $C$ for Case~A with $2\leq D\leq5$ for all the full-rank ML estimators that produced Fig.~\ref{fig:uSCA}. The measurement POM is, again, a set of $M=D^3$ square-root measurement outcomes for each $D$ that measures $N/M=500$. \emph{All} horizontal axes represent $C$, and vertical axes $S_{\mathcal{D}}$. The solid analytical curves are calculated using Eq.~\eqref{eq:caseA_dist}. All $S_{\mathcal{D}}$s are magnified---according to the magnification factors stated in the panels---so that all graphs and markers can be visibly co-plotted inside each panel.}
\end{figure}
\begin{figure}[t]
	\center
	\includegraphics[width=1\columnwidth]{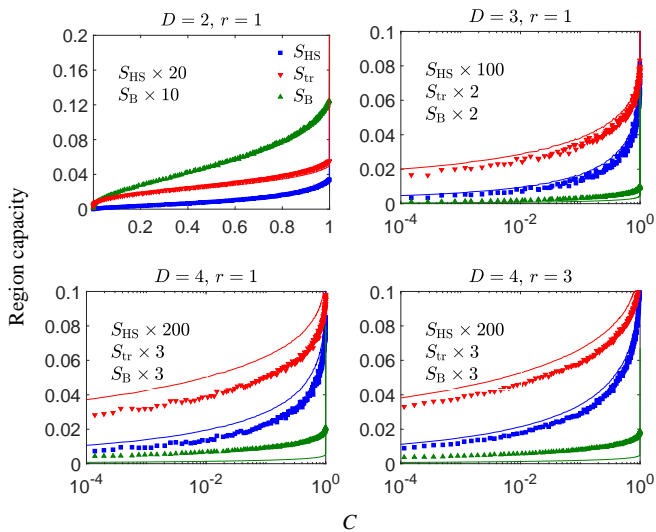}
	\caption{\label{fig:SCB}(Color Online) Graphs of the region capacity against $C$ for Case~B for all the rank-deficient ML estimators involved in Fig.~\ref{fig:uSCB}. \suggtwo{The solid curves, which originate from Eq.~\eqref{eq:caseB_dist}, still lie reasonably closely to the simulated markers for these low-dimensional examples.} All specifications otherwise follow those of Fig.~\ref{fig:SCA}. The geometrical difference between the actual rank-deficient CR boundary and a hyperplane manifests as deviations from theoretical predictions. High-rank ML estimators, nonetheless, generally gives a better theoretical predictions in contrast with low-rank estimators.}  
\end{figure}

In Fig.~\ref{fig:SCA}, both simulated data and theoretical curves of all three capacity functions $S_\textsc{hs}$, $S_\text{tr}$ and $S_\textsc{b}$ are plotted against the credibility $C$ for Case~A. In this case, there exists no other factors that could spoil the perfect hyperellipsoidal geometry of $\mathcal{R}_\lambda$. As such, the analytical curves fit almost perfectly with the simulated points. We note that even the average trace-class distance $S_\text{tr}$, which is approximated with \eqref{eq:tr2hs} through the theory of random matrices, performs very well relative to the simulated data points.

In Case~B, we can start to see discrepancies between theory and simulation from Fig.~\ref{fig:SCB} especially for larger $D$. Such deviations are inevitable as the hyperellipsoidal-cap estimation of the actual CR $\mathcal{R}_\lambda$ proposed in Sec.~\ref{subsec:hecap} introduces additional space outside $\mathcal{R}_0$ that is certainly not contained in $\mathcal{R}_\lambda$. More specifically, for very large $D$, if the rank-deficient ML estimator $\ML{\rho}$ is located at an extremely sharp state-space corner, which is labeled as the Type-I situation in Sec.~\ref{subsec:numcmpl}, this additional space would be exceedingly large relative to the physical CR, which incurs a proportionately large theory-simulation mismatch. On the other hand, if the rank-deficient $\ML{\rho}$ lies on a relatively flat part of the state-space boundary (the Type-II scenario), then this overestimated space, and hence the mismatch, would be much smaller. ML estimators of ranks 1 and 3, which are considered in Fig.~\ref{fig:SCB}, are prime examples of the respective Type-I and II situations. Regardless, the asymptotic formulas in Sec.~\ref{subsec:hecap} may still be used for an order-of-magnitude estimation of $S_{\mathcal{D}}$ and $C$.

\begin{figure}[t]
	\center
	\includegraphics[width=1\columnwidth]{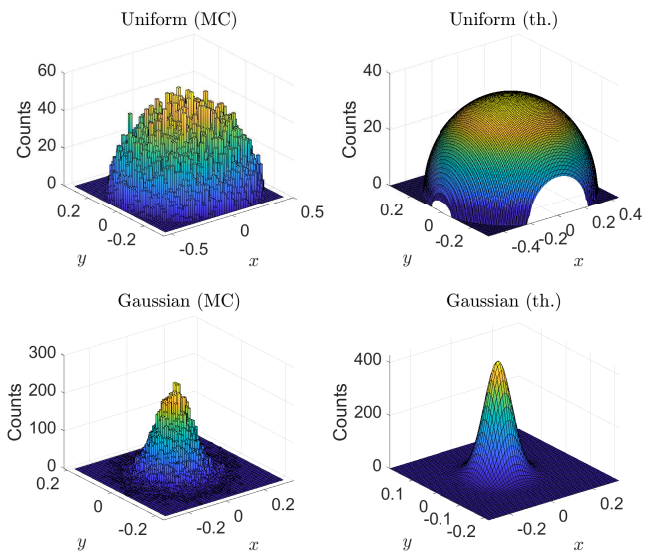}
	\caption{\sugg{\label{fig:calib_prior}(Color Online) Example comparisons between hit-and-run simulated and theoretical distributions made in Case~A, for $D=2$, random square-root measurement ($M=4$) and $N=50$. Their common coordinate system is centered at $\ML{\rvec{r}}$ and rotated in the frame of the error region. Both uniform and Gaussian (of covariance $10\,\FML$) prior distributions considered here are projected onto the error region, which is approximated as a hyperellipsoid for calculating the theoretical distributions (see Appendix~\ref{app:calib_prior} for their explicit probability-density expressions).}}  
\end{figure}

\sugg{We note that in-region sampling is not restricted to just the uniform distribution, so long as the MC method employed is sufficiently general. Such is the case for hit-and-run. For a calibration check of the general hit-and-run algorithm in Sec.~\ref{sec:numcomp}, we generate and compare both uniform and Gaussian distributions with their respective theoretically derived counterparts for the single-qubit case in Fig.~\ref{fig:calib_prior}.}
	
\sugg{Next, as a real demonstration, we consider another natural prior that is asymptotically conjugate to the likelihood function, that is the Gaussian form $p(\rvec{r})\propto\exp(-(\rvec{r}-\ML{\rvec{r}})\bm{\cdot}\FML\bm{\cdot}(\rvec{r}-\ML{\rvec{r}})/(2N))$ having a much broader covariance $N\invFML$ defined by the Fisher information for \emph{one copy}. This prior is a logical choice given that our knowledge about $\rvec{r}$ is updated to $\ML{\rvec{r}}$ after the measurement, which should be used as the most recent prior information for future Bayesian analyses. The corresponding marginal distribution needed to sample along line segments in hit-and-run is therefore the one-dimensional Gaussian distribution of mean $[\rvec{\mathrm{e}_v}\bm{\cdot}\FML\bm{\cdot}(\ML{\rvec{r}}-\rvec{r}_\text{ref})]/\rvec{\mathrm{e}_v}\bm{\cdot}\FML\bm{\cdot}\rvec{\mathrm{e}_v}$ and variance $1/\rvec{\mathrm{e}_v}\bm{\cdot}\FML\bm{\cdot}\rvec{\mathrm{e}_v}$.}

\begin{figure}[t]
	\center
	\includegraphics[width=1\columnwidth]{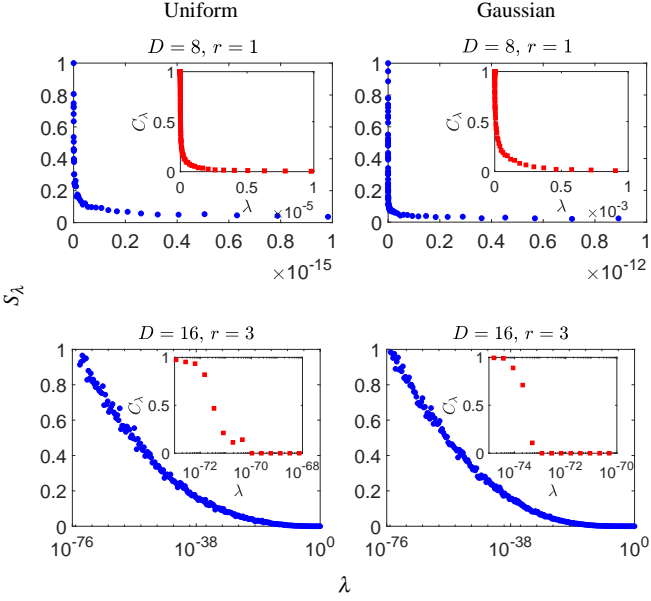}
	\caption{\label{fig:SC_8_16}(Color Online) \sugg{Graphs of $S_\lambda$ and $C_\lambda$ for $D=8$ and 16 under uniform and Gaussian prior distributions for Case B as explained in the main text. The measurement configurations are set to $N/M=5000$ and $N/M=50$ respectively for $D=8$ and 16, where $M=D^3$. All plot markers are computed with 20000 points generated during in-region sampling of $u_\lambda$ per $\lambda$. All Case-A credibility curves (not shown in this figure) match the theoretical results from Ref.~\cite{Teo:2018aa}. On the other hand, the Case-B theoretical curves for $C_\lambda$ can now be very different from the actual ones because of the complicated state-space boundary.}}
\end{figure}
\begin{figure}[t]
	\center
	\includegraphics[width=1\columnwidth]{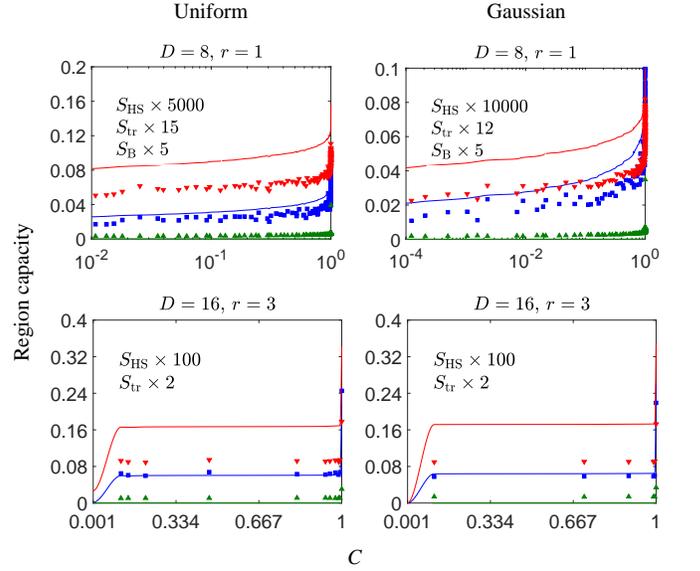}
	\caption{\label{fig:dist_8_16}(Color Online) Graphs of the region capacity against $C$ for all the rank-deficient ML estimators referred by Fig.~\ref{fig:SC_8_16}. \suggtwo{The theoretical curves for $S_\text{HS}$, $S_\text{tr}$ and $S_\text{B}$ are represented by the upper red, middle blue and bottom green solid curves respectively, whereas the simulated marker colors are as specified in Fig.~\ref{fig:SCB}.} The deviations from theoretical approximations for the region capacity, which is based on hyperplanar geometry, are apparently relatively more robust to high-dimensional state-space boundary features as compared to $S$ and $C$. More points are concentrated around large $C$.}  
\end{figure}

\subsection{Correlation properties of hit-and-run}

We recall that, at least under the uniform primitive prior $(\D\,\rvec{r}'')_\text{unif}$, hit-and-run converges efficiently to the correct uniform distribution in $O(d^2 R^2_\text{out}/R^2_\text{in})$ as discussed in Sec.~\ref{subsec:numcmpl}. Furthermore, it was argued~\cite{Lovasz:2006aa}, as a consequence of the above expression, that given an initial point that has a shortest distance $l$ from the boundary $\partial\mathcal{R}_\lambda$, hit-and-run eventually mixes sample points into the uniform distribution after $O(d^3 R^2_\text{out}/R^2_\text{in}\log(R_\text{out}/l))$ steps. 

\begin{figure*}[t]
	\center
	\includegraphics[width=2\columnwidth]{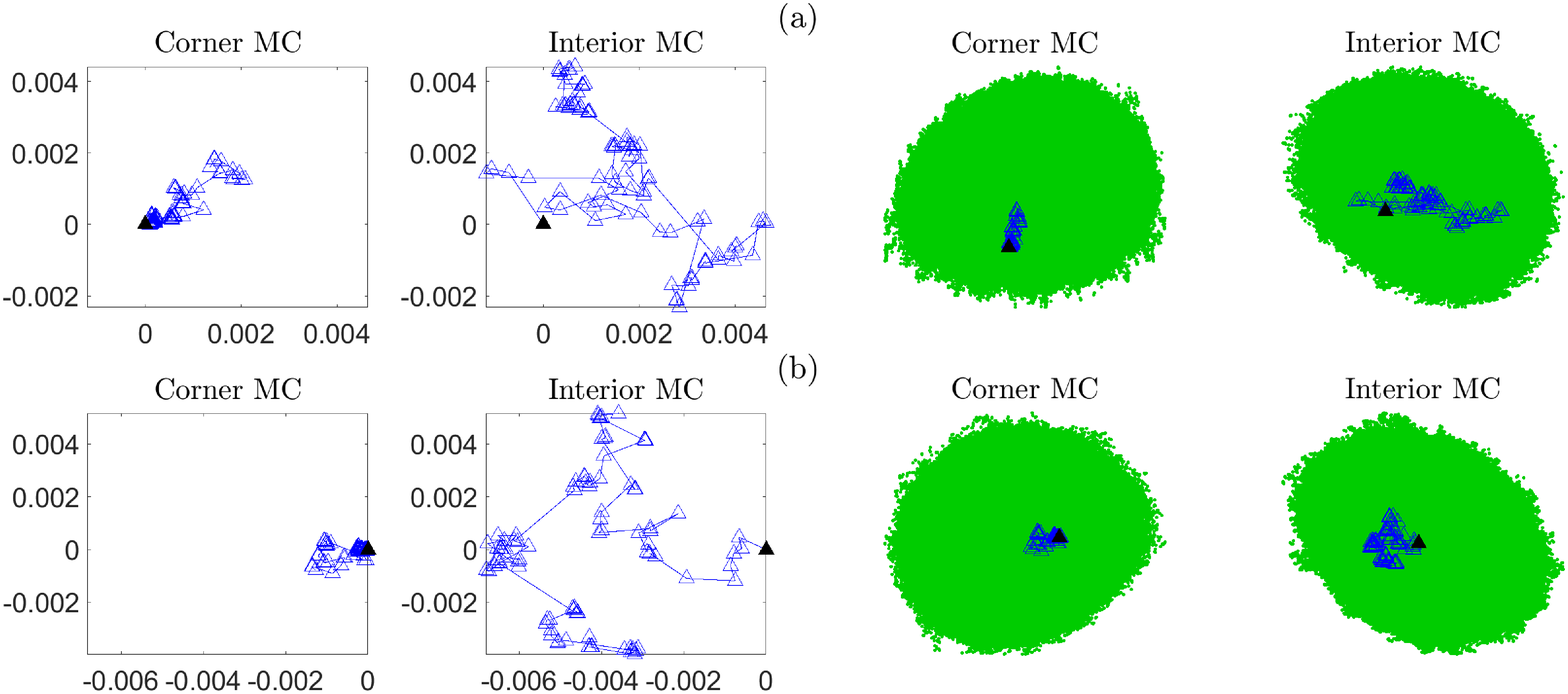}
	\caption{\label{fig:corr_unif}(Color Online) Case-B correlation strengths (translated to state-space hopping distances) of the first 100 hit-and-run MC-sampled points starting from both a corner of the CR (green area made up of 10 million uniformly sampled points) and an interior point, illustrated for two-qubit systems ($D=4$), a rank-one $\ML{\rho}$ obtained using $M=D^3$ and $N/M=500$, and a uniform primitive prior in $\rvec{r}$. The corner MC run starts (black shaded marker) from $\ML{\rho}$, and the interior run starts from a fully-mixed state inside the CR generated by averaging the first 1000 points of the hit-and-run algorithm beginning with $\ML{\rho}$. In properly scaled axes, corner MC shows a stronger correlation (shorter average hopping distances) than interior MC in the respective planes of (a) shortest and (b) longest average hopping distances, as the former encounters the region boundary much more frequently than the latter. This significantly limits the span of sampled points in the entire CR.}  
\end{figure*}
\begin{figure*}[t]
	\center
	\includegraphics[width=2\columnwidth]{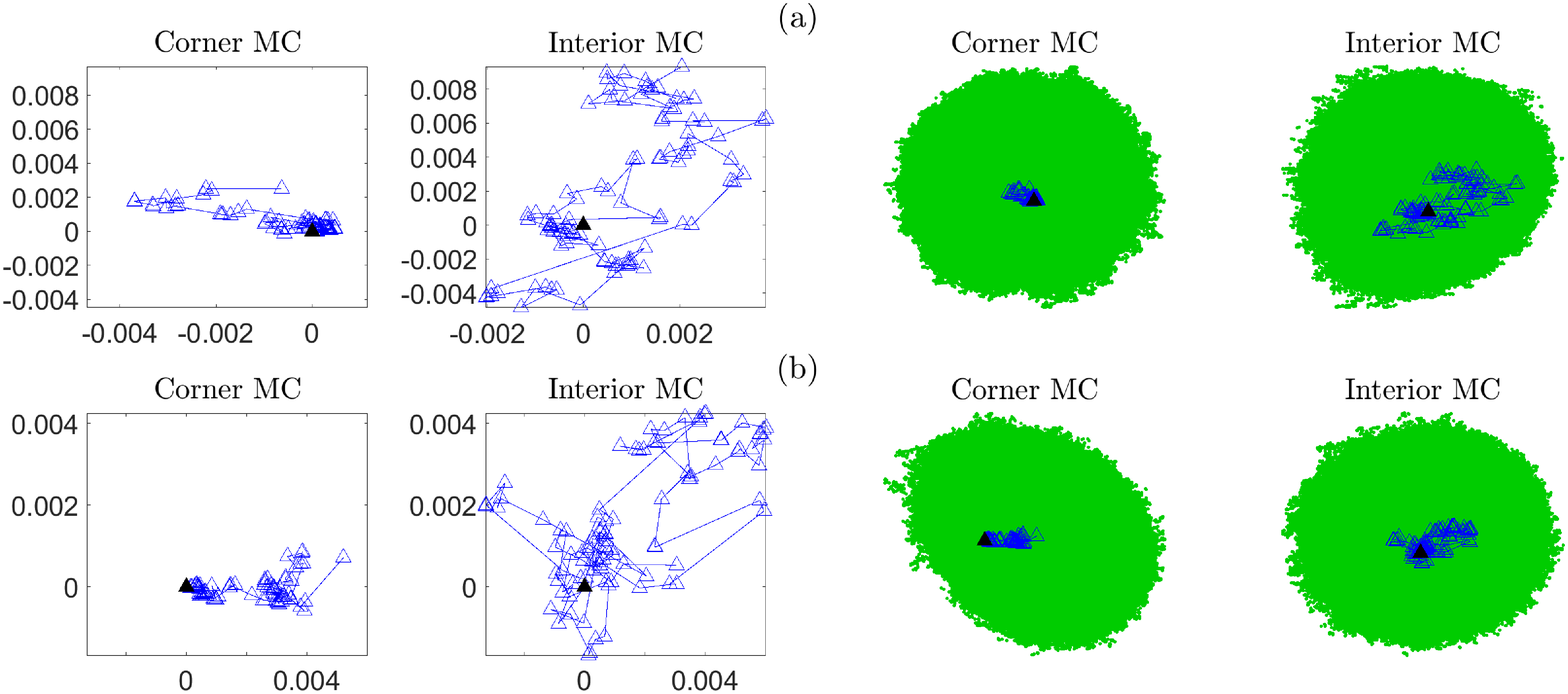}
	\caption{\sugg{\label{fig:corr_gauss}(Color Online) Case-B correlation strengths (translated to state-space hopping distances) of the first 100 hit-and-run MC-sampled points for the Gaussian prior distribution centered at $\ML{\rho}$. All figure specifications and conclusions are otherwise identical to those in Fig.~\ref{fig:corr_unif}.}}  
\end{figure*}

\sugg{This reveals a technical caveat for almost all Markov-chain random-walk algorithms: a random walk starting from a sharp corner point of a convex body requires very many steps to approach the stationary target distribution. It is generally much harder to scout the entire convex region from such a corner than from an interior point, since the Markov chain terminates as soon as a next admissible point is obtained, which is probabilistically near the initial corner point around which the admissible region is tight. Such \suggtwo{a} situation is essentially \emph{status quo} for high-dimensional state reconstruction where the state space $\mathcal{R}_0$ is filled with plenty of extremely sharp corners.}

\sugg{Doing hit-and-run from an interior point is therefore a primary objective for general CR analysis. Even without the full knowledge about the CR, it is still possible to numerically compute a point that is sufficiently interior for this purpose. The idea is to first find multiple random states on $\partial\mathcal{R}_\lambda\cap\mathrm{int}(\mathcal{R}_0)$, the boundary of $\mathcal{R}_\lambda$ in the interior of $\mathcal{R}_0$, and next average all these states to obtain an interior state of $\mathcal{R}_\lambda$. This is evidently equivalent to the minimization of the convex function $[(\rvec{x}'-\rvec{r}_\text{c})\bm{\cdot}\FML\bm{\cdot}(\rvec{x}'-\rvec{r}_\text{c})/(-2\log\lambda')-1]^2$ with respect to $\rho'\geq0$ for which $\rvec{x}'=\tr{\rho'\rvec{\Omega}}$ multiple times in the large-$N$ limit. Fortunately, this can be carried out extremely quickly by using the superfast accelerated projected gradient routine~\cite{Shang:2017sf} (see Appendix~\ref{app:bd_grad}).} 
	
\sugg{Figures~\ref{fig:corr_unif} and \ref{fig:corr_gauss} supply graphical visualization of the key sampling activities that goes on for Case~B with $D=4$, where a rank-one ML estimator is obtained. It is clear that starting hit-and-run from a corner point (namely the ML estimator, for instance) introduces small average hopping distances for subsequent Markov-chains. This can be interpreted as strong sample correlations that prevent wide coverage of the CR, contrary to performing hit-and-run starting from an interior point.}

\subsection{The constructions of plausible regions}

\sugg{The matter of inspecting $S_\lambda$ for a fixed $C_\lambda$, say 0.95, is rather subjective, for very often one requires experienced judgment to decide if such a value is sufficient for subsequent prediction tasks. As advocated in~\cite{Evans:2016aa,Al-Labadi:2018aa}, there exists a statistically meaningful treatment of the measured dataset $\mathbb{D}$ based on the concept of \emph{evidence}. \suggtwo{It} is thus fitting for us to end this article with a short review on how our in-region sampling technique studied here directly supports another interesting kind of Bayesian analysis.}

\sugg{By definition, we say that $\rvec{r}'$ is a \emph{plausible} candidate parameter for the true $\rvec{r}$ if there is indeed evidence in favor of this supposition. That is, its normalized posterior probability $L(\mathbb{D}|\rvec{r}')\,p(\rvec{r}')/L(\mathbb{D})$ after the measurement is larger than its prior probability $p(\rvec{r}')$ before this measurement was performed. Therefore, $\rvec{r}$ is a plausible parameter if the evidence supports the prior knowledge. Under this evidence-belief framework, one can construct another type of Bayesian region---the \emph{plausible region} (PR)---that contains \emph{all} plausible choices of $\rvec{r}$. This is really the CR $\mathcal{R}=\mathcal{R}_{\lambda=\lambda_{\rm{crit}}}$ characterized by the critical value~\cite{Li:2016da}
\begin{eqnarray}
\lambda_{\rm{crit}}=\int_0^1 d\lambda'\, S_{\lambda'}\,,
\label{eq:lbd_crit}
\end{eqnarray}
for which $L(\mathbb{D}|\rvec{r}\in\partial\mathcal{R}_{\lambda=\lambda_{\rm{crit}}})=L(\mathbb{D})$, or the CR that contains all plausible points and nothing else. This follows quickly from the following equality chain:
\begin{align}
L(\mathbb{D})=&\,\int(\D\,\rvec{r}')\,L(\mathbb{D}|\rvec{r}')=\int(\D\,\rvec{r}')\,\int^{L(\mathbb{D}|\rvec{r}')}_0\D x'\nonumber\\
=&\,L_\text{max}\int(\D\,\rvec{r}')\,\int^1_0\,\D\lambda'\,\eta\left(L(\mathbb{D}|\rvec{r}')-\lambda' L_\text{max}\right)\nonumber\\
=&\,L_\text{max}\int_0^1 d\lambda'\, S_{\lambda'}\,,
\end{align}
so that $L(\mathbb{D}|\rvec{r}\in\partial\mathcal{R}_{\lambda=\lambda_{\rm{crit}}})\equiv\lambda_{\rm{crit}}L_\text{max}=L(\mathbb{D})$ gives Eq.~\eqref{eq:lbd_crit}.}

\sugg{So, constructing a PR is nothing more than one additional step of computing $\lambda_\text{crit}$ after a CR construction. In our previous work~\cite{Teo:2018aa,Oh:2018aa}, we have supplied MC-less asymptotic approximations to the expression of $\lambda_\text{crit}$ for the uniform primitive prior. In the current context, it clearly follows that $\lambda_\text{crit}$ is also directly computable by simply doing a Riemann summation of the full $S_\lambda$ spectrum obtained through in-region sampling in accordance with Eq.~\eqref{eq:lbd_crit}.}

\section{Conclusions}

Quantum-state tomography is an important application of multidimensional parameter estimation. The construction of Bayesian credible regions for the reconstructed quantum states after tomography is, unfortunately, a highly nontrivial problem owing to the complex constraints inherited from the state space. Standard numerical recipe of first doing a Monte~Carlo sampling of the state space and next discarding points outside the credible region to compute its region qualities (size and credibility) quickly becomes infeasible when the dataset collected in an experiment is relatively big, as the corresponding credible region would be very small with respect to the state space. 

\sugg{In this article, we performed an extensive analysis of our recent in-region sampling technique that can construct credible regions that are usually very small in practice for reasonably high credibility values and large data samples. This technique computes credible-region qualities of a small credible region for any given prior distribution by inspecting how an appropriately chosen region-average quantity changes as the shape of the region varies. This procedure transforms the general credible-region construction into a sequence of direct region sampling followed by simple numerical solution to a single-variable differential equation. This results in no sample wastage since no points are discarded. The method of accelerated hit-and-run is one numerical scheme that can be used to compute region averages rather efficiently provided that sample correlation is properly mitigated with good Monte~Carlo starting points. One can also estimate its numerical complexity in the context of tomography despite the complicated state-space boundary.}

Furthermore, for highly complex quantum systems of extremely large dimensions, where all numerical methods eventually become practically infeasible, we derive a set of analytical formulas to perform approximate Bayesian error certification through the perspective of distance-induced \sugg{region capacity} measures that alternatively quantifies how large a credible region is. These formulas are now fully operational and further complement those for the conventional size function developed in previous works that require knowledge of the state-space volume.

\begin{acknowledgments}
The authors thank J.~Shang for fruitful discussions, and acknowledge financial support from the BK21 Plus Program (21A20131111123) funded by the Ministry of Education (MOE, Korea) and National Research Foundation of Korea (NRF), the framework of international cooperation program managed by the NRF (NRF-2018K2A9A1A06069933), and the Basic Science Research Program through the NRF funded by the Ministry of Education (No. 2018R1D1A1B07048633).
\end{acknowledgments}

\appendix

\section{The relationships between $S_\textsc{hs}$ and $S_\text{tr}$ in the large-$N$ (or $D$) limit}
\label{app:hsandtr}

Apart from $\mathcal{D}_\textsc{hs}$, all other measures have no direct analogs in the $\rvec{r}'$ parametrization. However in certain limits, all these measures have approximate relations with $\mathcal{D}_\textsc{hs}$.

We start with making an approximate connection between $S_\text{tr}$ and $S_\textsc{hs}$ by examining the Hermitian operator $\Delta\rho'=\rho'-\ML{\rho}$ ($\rho'\in\mathcal{R}$). In Case~A, the distribution of $\Delta\rho'$ in $\mathcal{R}$ has zero mean, $\overline{\Delta\rho'}^{\mathcal{R}}=0$. This is also approximately true for the Case~B situation when $N$ is sufficiently large such that $\mathcal{R}$ is small. Furthermore, the space of $\Delta\rho'$ is essentially a bounded set of Hermitian random operators. Here, we shall make the assumption that each matrix entry $\Delta\rho'_{jk}$ in the computational basis is an independent random complex number. Under this condition, the $\Delta\rho'$s form what is now known as a \emph{Wigner ensemble}~\cite{Wigner:1955aa,Wigner:1957aa,Pizzo:2013aa,Mehta:2004aa} with the second moment equal to $\overline{|\Delta\rho'_{jk}|^2}^{\mathcal{R}}=\tr{(\Delta\rho')^2}=S_{\textsc{hs}}$. Moreover, they are known to have an i.i.d. eigenvalue spectrum that follows the Wigner semicircle law 
\begin{align}
\sigma\left(\Delta\rho'/\sqrt{D}\right)\sim&\,\dfrac{1}{2\pi S_\textsc{hs}}\sqrt{4\,S_\textsc{hs}-x^2}\nonumber\\
&\qquad\quad\text{for }-2\sqrt{S_\textsc{hs}}\leq x\leq2\sqrt{S_\textsc{hs}}
\end{align}
in the large-$D$ limit. The trace-class distance $\mathcal{D}_\text{tr}$ can thus be calculated with the integral
\begin{align}
\mathcal{D}_\text{tr}\approx&\,\dfrac{\sqrt{D}}{2\pi S_\textsc{hs}}\int^{2\sqrt{S_\textsc{hs}}}_{-2\sqrt{S_\textsc{hs}}}\D\,x\,|x|\,\sqrt{4\,S_\textsc{hs}-x^2}=\dfrac{8\sqrt{D\,S_\textsc{hs}}}{3\pi}\,,
\end{align}
so that we end up with \eqref{eq:tr2hs}.

For Case~B, that $\overline{\Delta\rho'}^{\mathcal{R}}=0$ is obvious, but as we have no means of analytically estimate $\overline{\Delta\rho'}^{\mathcal{R}}$, we make a further approximation that as long as $\mathcal{R}$ is sufficiently small, the offset to $\overline{\Delta\rho'}^{\mathcal{R}}$ will proportionately be small, so that \eqref{eq:tr2hs} remains a reasonable asymptotic approximation.

\section{Fidelity in the large-$N$ limit}
\label{app:fid}
A Taylor expansion of $\mathcal{F}$ about $\ML{\rvec{r}}$, or
\begin{align}
\mathcal{F}\approx&\,1+\TP{(\rvec{r}'-\ML{\rvec{r}})}\,\dfrac{\partial\mathcal{F}_\textsc{ml}}{\partial\,\ML{\rvec{r}}}\nonumber\\
&\,\qquad\quad+\dfrac{1}{2}\TP{(\rvec{r}'-\ML{\rvec{r}})}\,\dfrac{\partial}{\partial\,\ML{\rvec{r}}}\dfrac{\partial\mathcal{F}_\textsc{ml}}{\partial\,\ML{\rvec{r}}}\,(\rvec{r}'-\ML{\rvec{r}})\,,
\end{align}
reveals the large-$N$ characteristics that is needed for analysis. The structure of \eqref{eq:fid}, however, demands the operator variation of $\sqrt{A}$ for a positive (semidefinite) $A$. An integral representation of $\sqrt{A}$ exists~\cite{Bini:2005aa} and can be written as
\begin{equation}
\sqrt{A}=\lim_{\epsilon\rightarrow 0^+}\int^\infty_0\dfrac{\D\,t}{\pi\sqrt{t}}\,\dfrac{A}{t+A+\epsilon}\,,
\label{eq:sqrtA}
\end{equation}
where the limit is understood to be applied at the very end of all calculations so that Eq.~\eqref{eq:sqrtA} is valid even for $A$ with zero eigenvalues.

The first-order variation of $\tr{\sqrt{A}}^2$ produces
\begin{align}
\updelta\,\tr{\sqrt{A}}^2=&\,2\,\tr{\sqrt{A}}\lim_{\epsilon\rightarrow 0}\int^\infty_0\dfrac{\D\,t}{\pi\sqrt{t}}\,\tr{\updelta\,\dfrac{A}{t+A+\epsilon}}\nonumber\\
=&\,2\,\tr{\sqrt{A}}\lim_{\epsilon\rightarrow 0}\int^\infty_0\dfrac{\D\,t}{\pi\sqrt{t}}\,\mathrm{tr}\bigg\{\updelta{A}\,\dfrac{1}{t+A+\epsilon}\nonumber\\
&\,\qquad\qquad\qquad\qquad-\dfrac{A}{t+A+\epsilon}\,\updelta A\,\dfrac{1}{t+A+\epsilon}\bigg\}\nonumber\\
=&\,2\,\tr{\sqrt{A}}\lim_{\epsilon\rightarrow 0}\tr{\updelta A\,\dfrac{A+2\,\epsilon}{2\,(A+\epsilon)^{3/2}}}\,.
\label{eq:dsqrtA}
\end{align}
In terms of $\mathcal{F}$, we substitute $A\equiv\ML{\rho}^{1/2}\,\rho'\,\ML{\rho}^{1/2}$, and evaluate the above result with $\rho'=\ML{\rho}$, or $A\rightarrow A_\textsc{ml}=\ML{\rho}^2$, then with $\updelta A_\textsc{ml}=\ML{\rho}^{1/2}\,\updelta\ML{\rvec{r}}\bm{\cdot}\rvec{\Omega}\,\ML{\rho}^{1/2}$,
\begin{align}
\dfrac{\partial\,\mathcal{F}_\textsc{ml}}{\partial\,\ML{\rvec{r}}}=&\,2\,\lim_{\epsilon\rightarrow 0}\tr{\ML{\rho}\,\dfrac{\ML{\rho}^2+2\,\epsilon}{2\,(\ML{\rho}^2+\epsilon)^{3/2}}\,\rvec{\Omega}}\,,
\end{align}
where we remind the Reader that $\mathrm{tr}$ acts on operators only, not on the vectorial character. For Case~B in which $\ML{\rho}=\sum^r_{j=1}\ket{\lambda_j}\lambda_j\bra{\lambda_j}$ is rank-deficient, we get, after taking the trace,
\begin{equation}
\dfrac{\partial\,\mathcal{F}_\textsc{ml}}{\partial\,\ML{\rvec{r}}}=\tr{P_r\,\rvec{\Omega}}\,,
\end{equation}
where $P_r=\sum^r_{j=1}\ket{\lambda_j}\bra{\lambda_j}$. It is then trivial to realize that this first-order derivative is zero for Case~A. Qualitatively, this confirms the fact that when $\ML{\rvec{r}}$ is an interior point, $\mathcal{F}$ has a local maximum at this point as it should, while a boundary estimator evaluates to a nonzero $\mathcal{F}$ slope.

Upon denoting $\rvec{W}_\textsc{ml}=\ML{\rho}^{1/2}\,\rvec{\Omega}\,\ML{\rho}^{1/2}$, the second-order variation follows from the second line of \eqref{eq:dsqrtA}:
\begin{align}
\updelta\,\dfrac{\partial\,\tr{\sqrt{A}}^2}{\partial\,\rvec{r}'}=&\,2\,\updelta\bigg[\tr{\sqrt{A}}\lim_{\epsilon\rightarrow 0}\int^\infty_0\dfrac{\D\,t}{\pi\sqrt{t}}\nonumber\\
&\,\qquad\qquad\times\tr{\rvec{W}_\textsc{ml}\dfrac{t+\epsilon}{(t+A+\epsilon)^2}}\bigg]\,.
\label{eq:ddsqrtA_def}
\end{align}
A product-rule dissociation of \eqref{eq:ddsqrtA_def} comprises a $\updelta\,\tr{\sqrt{A}}$ and 
\begin{align}
&\,\lim_{\epsilon\rightarrow 0}\int^\infty_0\dfrac{\D\,t}{\pi\sqrt{t}}\,\tr{\rvec{W}_\textsc{ml}\updelta\dfrac{t+\epsilon}{(t+A+\epsilon)^2}}\nonumber\\
=&\,-\lim_{\epsilon\rightarrow 0}\int^\infty_0\dfrac{\D\,t}{\pi\sqrt{t}}\,\mathrm{tr}\bigg\{\rvec{W}_\textsc{ml}\,\bigg[\dfrac{t+\epsilon}{t+A+\epsilon}\,\updelta A\,\dfrac{1}{(t+A+\epsilon)^2}\nonumber\\
&\,\qquad\qquad\qquad\qquad+\dfrac{t+\epsilon}{(t+A+\epsilon)^2}\,\updelta A\,\dfrac{1}{t+A+\epsilon}\bigg]\bigg\}\,.
\label{eq:ddsqrtA}
\end{align}
After evaluating the variation at $\rho'=\ML{\rho}$ and further undoing all integrations with the help of its spectral decomposition, Case~B yields
\begin{align}
\dfrac{\partial}{\partial\,\ML{\rvec{r}}}\dfrac{\partial\mathcal{F}_\textsc{ml}}{\partial\,\ML{\rvec{r}}}=&\,\dfrac{1}{2}\tr{P_r\,\rvec{\Omega}}\tr{P_r\,\TP{\rvec{\Omega}}}\nonumber\\
&\,\quad-\sum^r_{j=1}\sum^r_{k=1}\dfrac{\opinner{\lambda_j}{\rvec{\Omega}}{\lambda_k}\opinner{\lambda_k}{\TP{\rvec{\Omega}}}{\lambda_j}}{\lambda_j+\lambda_k}\,.
\end{align}
The counterpart expression for Case~A is immediate, of course.

\section{Hyperellipsoidal-cap averages}
\label{app:cap}

Under the uniform primitive prior, calculations of the hyperellipsoidal-cap integrals
\begin{align}
I_0=&\,\int(\D\,\rvec{r}'')_\text{cap}\,,\label{eq:I0}\\
\dyadic{I}_1=&\,\int(\D\,\rvec{r}'')_\text{cap}\,\,\rvec{r}''\,,\label{eq:I1}\\
\dyadic{I}_2=&\,\int(\D\,\rvec{r}'')_\text{cap}\,\,\rvec{r}''\TP{\rvec{r}''}\,,\label{eq:I2}
\end{align}
specified by the uniform cap-volume element $(\D\,\rvec{r}'')_\text{cap}=(\D\,\rvec{r}')_\text{unif}\,\eta(1-\rvec{r}''^2)\,\eta(a-\TP{\rvec{b}}\rvec{r}'')$ for $0\leq a\leq|\rvec{b}|$ and some column $\rvec{b}$, include systematic manipulations of the double Heaviside functions. One route to take exploits the following integral representation
\begin{equation}
\eta(x)=\int\dfrac{\D\,t}{2\pi\I}\,\dfrac{\E{\I x}}{t-\I\,\epsilon}
\label{eq:heavi_int}
\end{equation}
with the implicit limit $\epsilon\rightarrow0^+$. We then have, for \eqref{eq:I0},
\begin{align}
I_0=&\,\int\dfrac{\D\,t}{2\pi\I}\,\dfrac{\E{\I\, t}}{t-\I\,\epsilon}\int\dfrac{\D\,t'}{2\pi\I}\dfrac{\E{\I\,a\,t'}}{t'-\I\,\epsilon}\,\int(\D\,\rvec{r}'')_\text{unif}\,\E{-\I\,t\,\rvec{r}''^2-\I\,t'\,\TP{\rvec{b}}\rvec{r}''}\nonumber\\
=&\,\pi^{d/2}\int\dfrac{\D\,t}{2\pi}\,\dfrac{\E{\I\, t}}{(\I \,t)^{d/2+1}}\int\dfrac{\D\,t'}{2\pi}\dfrac{\E{\I\,a\,t'}}{\I \,t'}\,\E{\frac{\I}{4\,t}\,t'^2\rvec{b}^2}\,,
\end{align}
upon noting the well-known $d$-dimensional Gaussian integral result
\begin{equation}
\int(\D\,\rvec{r}'')_\text{unif}\,\E{-\TP{\rvec{r}''}\,\dyadic{A}\,\rvec{r}''+\TP{\rvec{c}}\rvec{r}''}=\dfrac{\pi^{d/2}}{\DET{\dyadic{A}}^{1/2}}\,\E{\frac{1}{4}\TP{\rvec{c}}\,\dyadic{A}\,\rvec{c}}
\label{eq:gauss_int}
\end{equation}
for any positive $\dyadic{A}$. Let us first do the $t'$ integration by invoking the useful transformation
\begin{equation}
\dfrac{1}{z^m}=\dfrac{1}{(m-1)!}\int^\infty_0\D\,y\,y^{m-1}\,\E{-z\,y}\quad\text{for }m>0\,:
\label{eq:gamma_int}
\end{equation}
\begin{align}
\int\dfrac{\D\,t'}{2\pi}\dfrac{\E{\I\,a\,t'}}{\I \,t'}\,\E{\frac{\I}{4\,t}\,t'^2\rvec{b}^2}=&\,\int^\infty_0\D\,y\,\int\dfrac{\D\,t'}{2\pi}\,\E{\frac{\I}{4\,t}\,t'^2\rvec{b}^2+\I\,(a-y)\,t'}\nonumber\\
=&\,\sqrt{\dfrac{\I\,t}{\pi\,\rvec{b}^2}}\,\int^\infty_0\D\,y\,\E{-\I\,t\,(a-y)^2/\rvec{b}^2}\,.
\end{align}
As a consequence,
\begin{align}
I_0=&\,\dfrac{\pi^{\frac{d-1}{2}}}{|\rvec{b}|}\int^\infty_0\D\,y\,\int\dfrac{\D\,t}{2\pi}\,\dfrac{\E{\I\, t}}{(\I \,t)^{\frac{d+1}{2}}}\,\E{-\I\,t\,(a-y)^2/\rvec{b}^2}\nonumber\\
=&\,\dfrac{\pi^{\frac{d-1}{2}}}{\left(\frac{d-1}{2}\right)!|\rvec{b}|}\int^\infty_0\D\,y\,\int^\infty_0\D\,y'\,y'^{\frac{d-1}{2}}\,\underbrace{\int\dfrac{\D\,t}{2\pi}\,\E{\I\,t\left[1-y'-(y-a)^2/\rvec{b}^2\right]}}_{\displaystyle=\delta(1-y'-(y-a)^2/\rvec{b}^2)}\nonumber\\
=&\,\dfrac{\pi^{\frac{d-1}{2}}}{\left(\frac{d-1}{2}\right)!|\rvec{b}|}\int^{a+|\rvec{b}|}_0\D\,y\,\left[1-(y-a)^2/\rvec{b}^2\right]^{\frac{d-1}{2}}\,.
\end{align}
The above integral in $y$ represents well-known special functions and to see this, we further perform the substitutions $\cos u=(y-a)/|\rvec{b}|$ and $l=a/b$:
\begin{align}
&\,\int^{a+|\rvec{b}|}_0\D\,y\,\left[1-(y-a)^2/\rvec{b}^2\right]^{\frac{d-1}{2}}\nonumber\\
=&\,|\rvec{b}|\,\int^{\cos^{-1}l}_0\D\,u\,(\sin u)^{d}\nonumber\\
=&\,|\rvec{b}|\,\BETA{\frac{1}{2},\frac{d+1}{2}}\mathrm{I}_{\frac{1-l}{2}}\left(\frac{d+1}{2},\frac{d+1}{2}\right)\,,
\end{align}
which is a product of the beta function and its normalized incomplete form
\begin{equation}
\mathrm{I}_{\,0\leq a\leq 1}\left(b,c\right)=\dfrac{1}{\BETA{b,c}}\int^a_0\D\,u\,\,u^b\,(1-u)^c\,.
\end{equation}
The final answer reads
\begin{equation}
I_0=V_d\,\mathrm{I}_{\frac{1-l}{2}}\left(\frac{d+1}{2},\frac{d+1}{2}\right)\,.
\end{equation}

For \eqref{eq:I1},
\begin{align}
\dyadic{I}_1=&\,\int\dfrac{\D\,t}{2\pi\I}\,\dfrac{\E{\I\, t}}{t-\I\,\epsilon}\int\dfrac{\D\,t'}{2\pi\I}\dfrac{\E{\I\,a\,t'}}{t'-\I\,\epsilon}\,\int(\D\,\rvec{r}'')_\text{unif}\,\rvec{r}''\,\E{-\I\,t\,\rvec{r}''^2-\I\,t'\,\TP{\rvec{b}}\rvec{r}''}\,,
\end{align}
where the $\rvec{r}''$ integration
\begin{align}
&\,\int(\D\,\rvec{r}'')_\text{unif}\,\rvec{r}''\,\E{-\I\,t\,\rvec{r}''^2-\I\,t'\,\TP{\rvec{b}}\rvec{r}''}\nonumber\\
=&\,-\dfrac{1}{\I\,t'}\,\dfrac{\partial}{\partial\,\rvec{b}'}\,\int(\D\,\rvec{r}'')_\text{unif}\,\E{-\I\,t\,\rvec{r}''^2-\I\,t'\,\TP{\rvec{b}}\rvec{r}''}\nonumber\\
=&\,-\I\,\dfrac{\pi^{d/2}\,t'}{2\,(\I\,t)^{d/2+1} }\,\rvec{b}\,\E{\frac{\I}{4\,t}\,t'^2\,\rvec{b}^2}
\end{align}
is simplified after a differentiation under the integral sign. Then
\begin{align}
\dyadic{I}_1=&\,-\dfrac{\pi^{d/2}}{2}\,\rvec{b}\int\dfrac{\D\,t}{2\pi}\,\dfrac{\E{\I\, t}}{(\I\,t)^{d/2+2}}\int\dfrac{\D\,t'}{2\pi}\,\E{\frac{\I}{4\,t}\,t'^2\,\rvec{b}^2+\I\,a\,t'}\nonumber\\
&\,-\dfrac{\pi^{\frac{d-1}{2}}}{2}\,\dfrac{\rvec{b}}{|\rvec{b}|}\,\int\dfrac{\D\,t}{2\pi}\,\dfrac{\E{\I\,t\,(1-l^2)}}{(\I\,t)^{\frac{d+3}{2}}}\,.
\end{align}
To simplify the $t$ integration, we again recall Eq.~\eqref{eq:gamma_int} to finally get
\begin{equation}
\dyadic{I}_1=-\dfrac{\pi^{\frac{d-1}{2}}}{2\left(\frac{d+1}{2}\right)!}\,\dfrac{\rvec{b}}{|\rvec{b}|}\,\left(1-l^2\right)^{\frac{d+1}{2}}\,,\quad l=\frac{a}{|\rvec{b}|}\,.
\label{eq:I1res}
\end{equation}
We can at least verify the $d=1$ for Eq.~\eqref{eq:I1res} after paying attention to the convention $\rvec{b}\rightarrow-b$, for $b\geq0$. This corresponds to the integral
\begin{align}
I_{1,d=1}=&\,\int\D\,r''\eta(1-r''^2)\,\eta(a+b\,r'')\,r''\nonumber\\
=&\,\int^1_{-1}\D\,r''\,\eta(a+b\,r'')\,r''\nonumber\\
=&\,\int^{-l}_{-1}\D\,r''\,r''=-\dfrac{1}{2}(1-l^2)\,.
\end{align}

By the same token, we may explore the dyadic integral $\dyadic{I}_2$ in \eqref{eq:I0} first with \eqref{eq:heavi_int} to obtain
\begin{align}
\dyadic{I}_2=&\,\int\dfrac{\D\,t}{2\pi\I}\,\dfrac{\E{\I\, t}}{t-\I\,\epsilon}\int\dfrac{\D\,t'}{2\pi\I}\dfrac{\E{\I\,a\,t'}}{t'-\I\,\epsilon}\nonumber\\
&\,\qquad\qquad\quad\int(\D\,\rvec{r}'')_\text{unif}\,\rvec{r}''\TP{\rvec{r}''}\,\E{-\I\,t\,\rvec{r}''^2-\I\,t'\,\TP{\rvec{b}}\rvec{r}''}\,,
\end{align}
where the dyadic $\rvec{r}''$ sub-integral
\begin{align}
&\,\int(\D\,\rvec{r}'')_\text{unif}\,\rvec{r}''\TP{\rvec{r}''}\,\E{-\I\,t\,\rvec{r}''^2-\I\,t'\,\TP{\rvec{b}}\rvec{r}''}\nonumber\\
=&\,\left.-\dfrac{1}{\I\,t}\,\dfrac{\updelta}{\updelta\,\dyadic{A}}\int(\D\,\rvec{r}'')_\text{unif}\,\E{-\I\,t\,\rvec{r}''^2-\I\,t'\,\TP{\rvec{b}}\rvec{r}''}\right|_{\dyadic{A}=\dyadic{1}}\nonumber\\
=&\,\left.-\dfrac{\pi^{d/2}}{(\I\,t)^{d/2+1}}\,\dfrac{\updelta}{\updelta\,\dyadic{A}}\,\dfrac{1}{\DET{\dyadic{A}}^{1/2}}\,\E{\frac{\I\,t'^2}{4\,t}\,\TP{\rvec{b}}\,\dyadic{A}\,\rvec{b}}\right|_{\dyadic{A}=\dyadic{1}}
\end{align}
after an application of Eq.~\eqref{eq:gauss_int} and a dyadic differentiation under the integral sign this time. 

This time, we choose to perform the $t$ and $t'$ integrals before taking the derivative, inasmuch as
\begin{equation}
\dyadic{I}_2=\left.-\dfrac{\updelta}{\updelta\,\dyadic{A}}\,\dfrac{\pi^{d/2}}{\DET{\dyadic{A}}^{1/2}}\int\dfrac{\D\,t}{2\pi}\,\dfrac{\E{\I\, t}}{(\I\,t)^{d/2+2}}\int\dfrac{\D\,t'}{2\pi}\dfrac{\E{\I\,a\,t'}}{\I\,t'}\,\E{\frac{\I\,t'^2}{4\,t}\,\TP{\rvec{b}}\,\dyadic{A}\,\rvec{b}}\right|_{\dyadic{A}=\dyadic{1}}\,,
\label{eq:I2_inter}
\end{equation}
where the usage of \eqref{eq:gamma_int} evaluates the $t'$ integral
\begin{align}
&\,\int\dfrac{\D\,t'}{2\pi}\dfrac{\E{\I\,a\,t'}}{\I\,t'}\,\E{\frac{\I\,t'^2}{4\,t}\,\TP{\rvec{b}}\,\dyadic{A}\,\rvec{b}}\nonumber\\
=&\,\int^\infty_0\D\,y\,\int\dfrac{\D\,t'}{2\pi}\,\E{\frac{\I\,t'^2}{4\,t}\,\TP{\rvec{b}}\,\dyadic{A}\,\rvec{b}+\I\,t'\,(a-y)}\nonumber\\
=&\,\sqrt{\dfrac{\I\,t}{\pi\,\TP{\rvec{b}}\,\dyadic{A}\,\rvec{b}}}\int^\infty_0\D\,y\,\exp\!\left(-\I\,\dfrac{(y-a)^2t}{\TP{\rvec{b}}\,\dyadic{A}\,\rvec{b}}\right)
\end{align}
into another Gaussian integral. Its convenient feature becomes clear when substituted back into \eqref{eq:I2_inter}:
\begin{align}
\dyadic{I}_2=&\,-\dfrac{\updelta}{\updelta\,\dyadic{A}}\,\dfrac{\pi^{\frac{d-1}{2}}}{\sqrt{\DET{\dyadic{A}}\TP{\rvec{b}}\,\dyadic{A}\,\rvec{b}}}\nonumber\\
&\,\times\int\dfrac{\D\,t}{2\pi}\,\dfrac{\E{\I\, t}}{(\I\,t)^{\frac{d+3}{2}}}\int^\infty_0\D\,y\,\exp\!\left(-\I\,\dfrac{(y-a)^2t}{\TP{\rvec{b}}\,\dyadic{A}\,\rvec{b}}\right)\Bigg|_{\dyadic{A}=\dyadic{1}}\nonumber\\
=&\,-\dfrac{\updelta}{\updelta\,\dyadic{A}}\,\dfrac{\pi^{\frac{d-1}{2}}}{\sqrt{\DET{\dyadic{A}}\TP{\rvec{b}}\,\dyadic{A}\,\rvec{b}}}\vphantom{\Bigg|^{\Bigg|}}\nonumber\\
&\,\times\dfrac{1}{\left(\frac{d+1}{2}\right)!}\int^\infty_0\D\,y\int^\infty_0\D\,y'\,y'^{\frac{d+1}{2}}\nonumber\\
&\,\times\underbrace{\int\dfrac{\D\,t}{2\pi}\,\exp\!\left(-\I\,t\left(1-y'-\dfrac{(y-a)^2t}{\TP{\rvec{b}}\,\dyadic{A}\,\rvec{b}}\right)\right)}_{\displaystyle=\delta\!\left(1-y'-\dfrac{(y-a)^2t}{\TP{\rvec{b}}\,\dyadic{A}\,\rvec{b}}\right)}\Bigg|_{\dyadic{A}=\dyadic{1}}\nonumber\\
=&\,-\dfrac{\pi^{\frac{d-1}{2}}}{\left(\frac{d+1}{2}\right)!}\,\dfrac{\updelta}{\updelta\,\dyadic{A}}\,\dfrac{1}{\sqrt{\DET{\dyadic{A}}\TP{\rvec{b}}\,\dyadic{A}\,\rvec{b}}}\nonumber\\
&\,\times\int^{a+\sqrt{\TP{\rvec{b}}\,\dyadic{A}\,\rvec{b}}}_0\D\,y\left[1-\dfrac{(y-a)^2}{\TP{\rvec{b}}\,\dyadic{A}\,\rvec{b}}\right]^{\frac{d+1}{2}}\nonumber\\
=&\,\left.-\dfrac{\pi^{\frac{d-1}{2}}}{\left(\frac{d+1}{2}\right)!}\,\dfrac{\updelta}{\updelta\,\dyadic{A}}\,\dfrac{1}{\DET{\dyadic{A}}^{1/2}}\int^{\cos^{-1} l_{\dyadic{A}}}_0\D\,u\,(\sin u)^{d+2}\right|_{\dyadic{A}=\dyadic{1}}\,.
\end{align}

The end of the tunnel becomes visible after a product-rule differentiation carried out with the basic dyadic identities
\begin{align}
\updelta\dyadic{A}^{-1}=&\,-\dyadic{A}^{-1}\updelta\dyadic{A}\,\,\dyadic{A}^{-1}\nonumber\\
\updelta\DET{\dyadic{A}}=&\,\DET{\dyadic{A}}\Tr{\dyadic{A}^{-1}\updelta\dyadic{A}}\,,
\end{align}
after which we end up with the final answer
\begin{align}
\dyadic{I}_2=&\,\dfrac{\pi^{\frac{d-1}{2}}}{2\left(\frac{d+1}{2}\right)!}\bigg[\BETA{\frac{1}{2},\frac{d+3}{2}}\mathrm{I}_{\frac{1-l}{2}}\!\left(\frac{d+3}{2},\frac{d+3}{2}\right)\dyadic{1}\nonumber\\
&\,+l\left(1-l^2\right)^{\frac{d+1}{2}}\dfrac{\rvec{b}\,\TP{\rvec{b}}}{\rvec{b}^2}\bigg]\,.
\label{eq:I2res}
\end{align}
The 1D special case can again be extracted from Eq.~\eqref{eq:I2res},
\begin{align}
\dyadic{I}_2\Big|_{d=1}=&\,\dfrac{1}{2}\left[\BETA{\frac{1}{2},2}\mathrm{I}_{\frac{1-l}{2}}\!\left(2,2\right)+l(1-l^2)\right]\nonumber\\
=&\,\dfrac{1}{2}\left[8\int^{\frac{1-l}{2}}_0\D\,u\,u(1-u)+l(1-l^2)\right]\nonumber\\
=&\,\dfrac{1}{2}\left[8\left(\frac{l^3}{24}-\frac{l}{8}+\frac{1}{12}\right)+l(1-l^2)\right]\nonumber\\
=&\,\frac{1}{3}\left(1-l^3\right)
\end{align}
and compared with the direct calculation
\begin{align}
I_{2,d=1}=&\,\int\D\,r''\eta(1-r''^2)\,\eta(a+b\,r'')\,r''^2\nonumber\\
=&\,\int^1_{-1}\D\,r''\,\eta(a+b\,r'')\,r''^2\nonumber\\
=&\,\int^{-l}_{-1}\D\,r''\,r''^2=\dfrac{1}{3}(1-l^3)\,.
\end{align}

\section{Projection of uniform and Gaussian distributions onto a hyperellipsoid}
\label{app:calib_prior}

\sugg{One may begin with a rotated coordinate system (centered at $\ML{\rvec{r}}$) that diagonalizes the Fisher information $\FML$, so that the projected uniform distribution onto the error region in the large-$N$ approximation is calculated from the integral
\begin{equation}
p_\text{unif}\propto\int\D z\,\eta(1-a x^2-b y^2-c z^2)
\end{equation}
for the eigenvalues $a$, $b$, and $c$ of $\FML/(-2\log\lambda)$. The preceding exercises of Appendix~\ref{app:cap} swiftly gives $p_\text{unif}\propto\sqrt{1-a x^2-b y^2}$.}

\sugg{The Gaussian distribution, with covariance chosen to be proportional to $\FML$ that possesses the eigenvalues $a'$, $b'$ and $c'$, is given by
\begin{equation}
p_\text{gauss}\propto\int\D z\,\eta(1-a x^2-b y^2-c z^2)\,\E{-a'x^2-b'y^2-c'z^2}\,.
\end{equation}
This can be simplified to $p_\text{gauss}\propto\E{-a'x^2-b'y^2}\,\gamma(1/2,c'(1-ax^2-by^2)/c)$ in terms of the lower incomplete Gamma function $\gamma(\cdot,\cdot)$ using again results from Appendix~\ref{app:cap}.}

\section{Gradient optimization for obtaining an error-region interior point}
\label{app:bd_grad}

\sugg{To acquire an interior point of a CR for state tomography, which is essentially a Hilbert subspace, it is sufficient to generate very many ($>D$) region boundary points and take the average of these points. In the limit of large $N$, we may approximate the inner boundary of the region as part of the hyperellipsoid described by $(\rvec{x}'-\rvec{r}_\text{c})\bm{\cdot}\dyadic{A}\bm{\cdot}(\rvec{x}'-\rvec{r}_\text{c})\leq1$.}

\sugg{If we perform a variation on the relevant function 
\begin{equation}
f=[(\rvec{x}'-\rvec{r}_\text{c})\bm{\cdot}\dyadic{A}\bm{\cdot}(\rvec{x}'-\rvec{r}_\text{c})-1]^2\,,
\label{eq:f}
\end{equation}
where $\dyadic{A}=\FML/(-2\log\lambda')$, we get
\begin{equation}\\
\updelta f=4\,[(\rvec{x}'-\rvec{r}_\text{c})\bm{\cdot}\dyadic{A}\bm{\cdot}(\rvec{x}'-\rvec{r}_\text{c})-1]\,\updelta\,\rvec{x}'\bm{\cdot}\dyadic{A}\bm{\cdot}(\rvec{x}'-\rvec{r}_\text{c})\,.
\end{equation}
A gradient method, such as the accelerated projected gradient method, requires the definition of the (operator) gradient defined by $\updelta f/\updelta \rho'$, which requires the connection $\updelta\,\rvec{x}'=\tr{\updelta \rho'\,\dyadic{\Omega}}$. Naturally then, we must acquire the resulting operator
\begin{equation}
\label{eq:df}
\dfrac{\updelta f}{\updelta \rho'}=4\,[(\rvec{x}'-\rvec{r}_\text{c})\bm{\cdot}\dyadic{A}\bm{\cdot}(\rvec{x}'-\rvec{r}_\text{c})-1]\,\,\dyadic{\Omega}\bm{\cdot}\dyadic{A}\bm{\cdot}(\rvec{x}'-\rvec{r}_\text{c})\,,
\end{equation}
where the dot products operate only on the vectorial character only.
}

\sugg{The mechanisms that drive the accelerated projected gradient search algorithm are beyond the scope of this article. Instead we provide a simple manual to immediately modify and use the open-source MATLAB code file \texttt{qse\us apg.m} that is available on \url{https://github.com/qMLE/qMLE}. For this purpose, we note the three important variables \texttt{fval\us varrho}, \texttt{fval\us new} and \texttt{gradient}, which stores the function values of $f$ evaluated with the \texttt{varrho} and \texttt{rho\us new} variables, as well as the gradient operator $\updelta f/\updelta \rho'$ evaluated with \texttt{varrho}. In order to minimize $f$ with \texttt{qse\us apg.m}, one may simply overwrite the existing functional expressions [namely \texttt{-f'.*log(probs\us\ldots)} and \texttt{-qmt(\ldots)}] for the three variables with the ones in Eqs.~\eqref{eq:f} and \eqref{eq:df}. By our numerical experience with this minimization task, it is advisable to set the parameters \texttt{defaults.threshold\us fval} and \texttt{defaults.imax} respectively to \texttt{eps} and $>10^{-8}$ for better accuracies.}

%



\end{document}